\documentclass[preprint,11pt]{JHEP3} % 10pt is ignored!

\JHEPspecialurl{http://jhep.sissa.it/JOURNAL/JHEP3.tar.gz}

\usepackage{epsfig,multicol,amsmath}

\usepackage{rotating}
\usepackage{cite}

\def\beq{\begin{equation}}
\def\eeq{\end{equation}}
\def\bea{\begin{eqnarray}}
\def\eea{\end{eqnarray}}

\def\eq#1{{Eq.~(\ref{#1})}}
\def\fig#1{{Fig.~\ref{#1}}}

\newcommand{\Lb}{\left(}
\newcommand{\Rb}{\right)}
\parskip=\itemsep               %?

\def\thefootnote{\fnsymbol{footnote}}

\title{ Survival Probabilities for High Mass Diffraction}
\author{\Large
\,E. Gotsman\thanks{Email: gotsman@post.tau.ac.il}\,, \,A.
Kormilitzin\thanks{Email: andreyk1@post.tau.ac.il}\,, \,E.
Levin\thanks{Email: leving@post.tau.ac.il,levin@mail.desy.de} \,and
\,U. Maor \thanks{Email: maor@post.tau.ac.il}
\\
Department of Particle Physics, School of Physics and Astronomy\\
Raymond and Beverly Sackler Faculty of Exact Science\\
Tel Aviv University, Tel Aviv, 69978, Israel}

\abstract {
Based on the calculation of  survival probabilities,
we suggest a procedure to assess  the value of $G_{3P}$,
the triple Pomeron 'bare' coupling constant, by comparing the large rapidity
 gap
single high mass diffraction data in proton-proton scattering
and $J/\Psi$ photo and DIS production.
For $p$-$p$ scattering the calculation  in a three amplitude rescattering
eikonal model,  predicts the survival probability to be
an order of magnitude smaller than for the two amplitude case.
The calculation of the survival probabilities  for photo and DIS $J/\Psi$
production are made  in a dedicated model.
In this process we show that, even though its
survival probability is considerably larger than in $p$-$p$ scattering, its value
is
below unity and cannot be neglected in the data analysis.
We argue that, regardless of the uncertainties in the suggested
procedures, the outcome is important,  both with regards to a realistic estimate of
$G_{3P}$, and the survival probabilities relevant to LHC experiments.
}
\dedicated{PACS: 13.85-t, 13.85.Hd, 11.55.-m, 11.55.Bq}
\preprint{TAUP -2846-07\\
\today}
\begin{document}
\def\thefootnote{\arabic{footnote}}
%%%%%%%%%%%%%%%%%%%%%%%%%%%%%%%%%%%%%%%%%%%%%%%%%%%%%%%%%%%%%%%%%%%%%%%%
\section{Introduction}
\label{sec:Int}
%%%%%%%%%%%%%%%%%%%%%%%%%%%%%%%%%%%%%%%%%%%%%%%%%%%%%%%%%%%%%%%%%%%%%%%%
A  large rapidity gap (LRG) process is defined as one where no
hadrons are produced in a sufficiently large rapidity interval.
Diffractive LRG
are assumed to be produced by the exchange of a color singlet object
with quantum numbers of the vacuum, which we will refer to as the Pomeron.
We wish to estimate the probability that a LRG, which
occurs in diffractive events,
survives rescattering effects which populate the gap with
secondary particles coming from the underlying event.

At high energies, elastic and inelastic diffractive processes account
for about $40\%$ of the total
$p$-$p$ ($\bar{p}$-$p$) cross section.
We would like to remind the reader that:
\begin{enumerate}
\item  The small $t$ behavior of the scattering amplitude is determined, mostly,
by the large impact parameter $b$ values.
\item The survival probability $<|S|^{2}>$ (denoted $S^{2}$)
of a diffractive LRG is obtained from a
normalized convolution of the
b-space diffractive amplitude squared
and $e^{-\Omega(s,b)}$.
${\Omega(s,b)}$ is the optical density, also called  opacity.
Consequently, $S^{2}$ decreases with increasing energy due to
the growth with energy of the interaction input opacity.
\item  $S^{2}$ is not only dependent on the probability of the initial
state to survive, but is also sensitive  to the spatial distribution
of the
partons inside the incoming hadrons and, thus, on the dynamics of the whole
diffractive part of the scattering matrix.
\item  $S^{2}$, at a given energy, is not universal. It depends on the particular
diffractive subprocess, as well as the kinematic configurations. It also depends on
the nature of the color singlet (P, W/Z or $\gamma$) exchange which is
responsible for the LRG.
\end{enumerate}

Historically, both Dokshitzer et al.\cite{Dok} and  Bjorken\cite{Bj}
suggested  utilizing  a  LRG as a signature for Higgs
production originating from a W-W fusion sub-process,
in hadron-hadron collisions.
It turns out that  LRG processes give a unique opportunity to
measure the high energy asymptotic behavior of the amplitudes at short
distances, where one can calculate the amplitudes using methods developed
for
perturbative QCD (pQCD).
Consider a typical LRG process - the production of two jets
with large transverse momenta
$\vec{p}_{t1}\,\,\approx\,\,-\vec{p}_{t2}\,\,\gg\,\,\mu $, with a LRG
between the two jets.
$\mu$ is a typical mass scale of the soft interactions.
\bea \label{I1}
& p (1)\,\,+\,\,p(2)\,\,\longrightarrow\,\,
M_1[ hadrons\,\,+\,\,jet_1(y_1,p_{t1})]& \\
&
+\,\,LRG[ \Delta y = |y_1 - y_2|]
\,\,+\,\,M_2[ hadrons\,\,+\,\,jet_2(y_2,p_{t2})].& \nonumber
\eea
$y_1$ and $y_2$ are the rapidities of the jets and
$\Delta y \,\,\gg\,\,1$. The production of two hard jets with a LRG is
initiated by the exchange of a hard Pomeron.
We define $F_s$ to be the ratio between the
cross section due to the above Pomeron exchange, and the inclusive inelastic
cross section with the same final state generated by gluon exchange.
In QCD we do not expect this ratio to decrease as a function of the
rapidity gap $\Delta y$. For a BFKL Pomeron\cite{BFKL},
we expect an increase once  $\Delta y \,\gg\,\,1$. Using
a  simple QCD model for the Pomeron, in which it is approximated
by two gluon exchange\cite{LN}, Bjorken\cite{Bj} gave the
first estimate for $F_s \,\approx\,\,0.15 $,
which is unexpectedly large.

As noted by Bjorken\cite{Bj} and GLM\cite{GLM1},
one does not measure $F_s$ directly in a LRG experiment.
The experimentally measured ratio between the number of events with a LRG, and
the number of similar events without a LRG
is not equal to $F_s$, but, has to be modified by an extra suppression factor
which we call the LRG survival probability,
\beq \label{I2}
f_{gap} = <\mid S \mid^2>\times F_{s}\,.
\eeq
The appearance of $S^2$ in \eq{I2} has a very
simple physical interpretation. It is the  probability that the
LRG due to Pomeron exchange, will not be filled by the
produced particles (partons and/or hadrons) from the rescattering
of the spectator partons, or from the emission of
bremsstrahlung gluons coming from the partons, or the hard Pomeron,
taking part in the hard interaction.
\beq \label{I3}
<\mid S \mid^2>\,\,=
\,\,<\mid S_{bremsstrahlung} (\Delta y\,=\,|\,y_1\,-\,y_2\,|)\mid^2> \times
<\mid S_{spectator}(\,s\,) \mid^2>\,,
\eeq
where $s$ denotes the total c.m. energy squared.
\begin{itemize}
\item\,\,\, $S_{bremsstrahlung}^2(\Delta y)$
can be calculated in pQCD\cite{BREM}.
It depends on the kinematics of each specific process,
and on the value of the LRG . This suppression is commonly included
in the calculation of the hard LRG sub process.
\item\,\,\,To calculate $S_{spectator}^2(s)$ we need to
find the probability that all partons with rapidity $y_i\,>\,y_1$
in the first hadron, and all partons with
$y_j\,<\,y_2$ in the second hadron, do not interact inelastically
and, hence, do not produce additional hadrons within the LRG interval. This
is a  difficult problem, since  not only partons at short
distances contribute to such a calculation, but also partons at long
distances,   for which the pQCD approach is not valid.
Many attempts have been made to estimate
$S_{spectator}^2$\cite{Bj,GLM1,ELSP,FLET,RR,GLMSM,GLMSP2},
but a unique solution, has still not been found.
\end{itemize}

An obvious check of the above is to compare the calculated values of
$S^2_{spectator}(s)$ obtained in different models for different reactions.
The Durham group\cite{DG3P}  recently suggested a very interesting procedure,
proposing to extract $G_{3P}$, the triple Pomeron vertex coupling,
utilizing the measurement of large mass diffraction
dissociation in the reaction
\beq \label{REACT}
\gamma^*(Q^2,x_{Bj})\,+\,p\,\Longrightarrow\,
J/\Psi\,+ \,[LRG]\,+\,X(M^2 \gg\ m_p).
\eeq

The cross section of this process can be described by
the Mueller diagram (\fig{3pjdia}).
It is initiated by the charm component of the photon which has a small
absorptive cross section, since its interaction stems from
short distances ($r \propto 1/m_c$, where $m_c$ is the mass of the charm quark).
Thus, the probability for additional rescatterings (\fig{3pspjdia})
is relatively small, resulting in a high survival probability.
This is to be compared with
the corresponding high mass diffraction in an hadronic
$p$-$p$ ($\bar{p}$-$p$) reaction (\fig{3pdia}), for which we expect the
rescatterings (\fig{3pspdia}) to be significant, resulting in a small
survival probability\cite{3PSC}.
It is, therefore, very probable that present
extractions\cite{3PP,3POLD} of the 3P coupling are underestimated.
Comparing the values of $G_{3P}$ obtained
in the above two channels, taking into account their (different)
survival probabilities, leads to a more reliable
measure of the 3P coupling, and provides a check of
the various theoretical estimates of the survival probabilities.
%%%%%%%%%%%%%%%%%%%%%%%%%%%%%%%%%%%%%%%%%%%%%%%%%%%%%%%%%%%%%%%%%%%%%%%%
\DOUBLEFIGURE[ht]{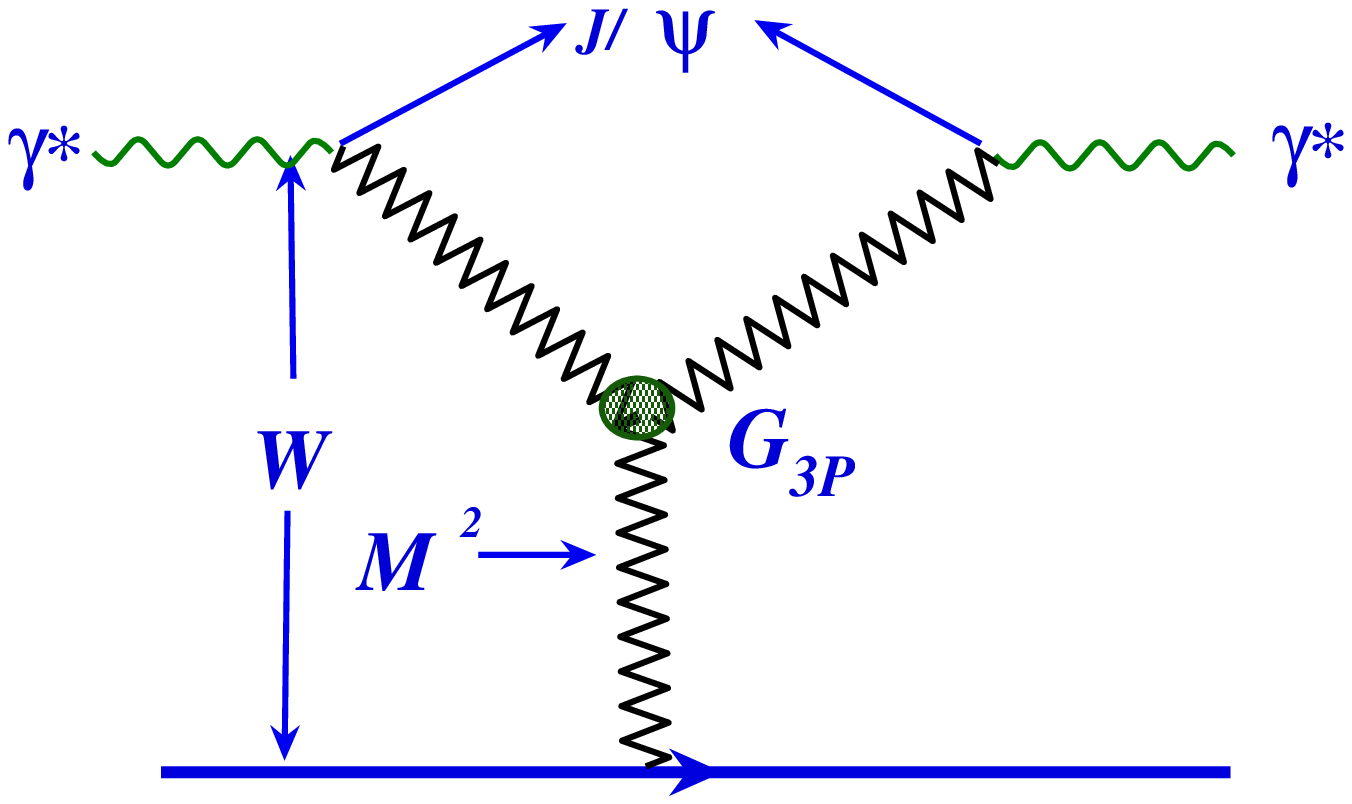,width=70mm,height=40mm}
{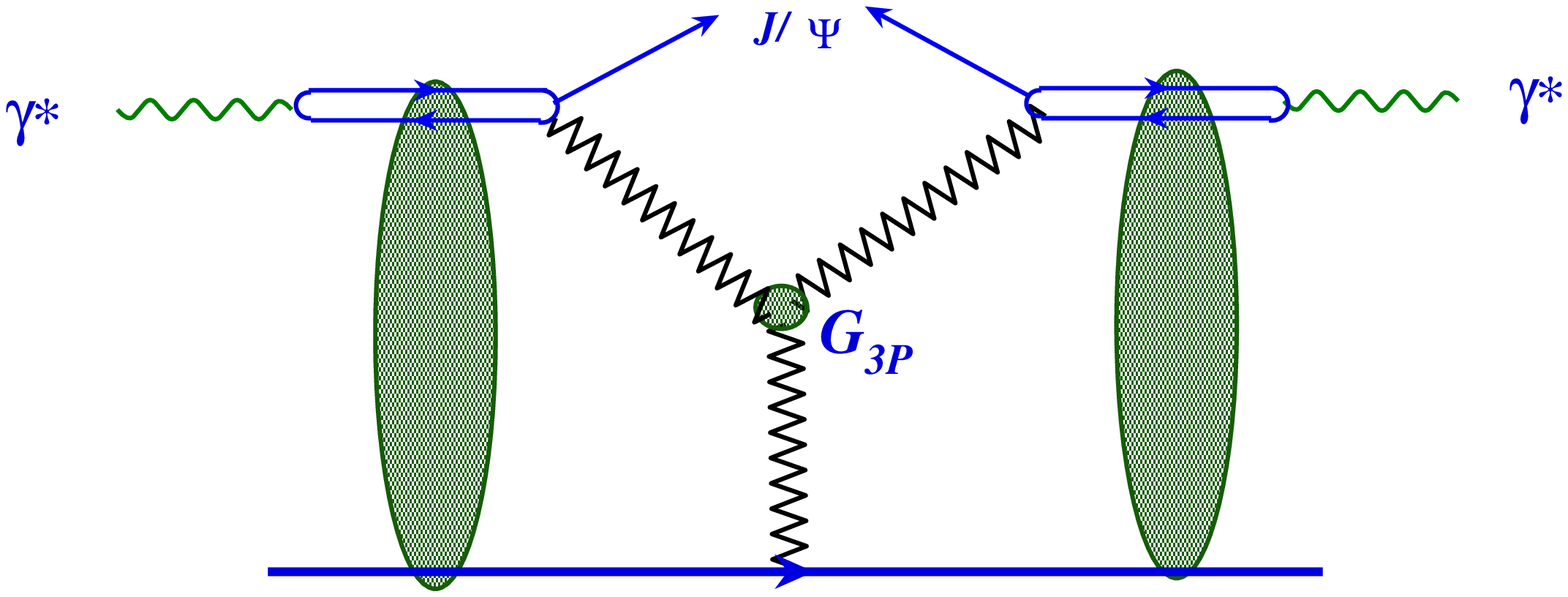,width=80mm,height=40mm}
{The general diagram for diffractive production of large masses
in $\gamma^*$-$p$ collisions at high energy.
Pomerons are denoted by the zigzag lines. The vertical lower Pomeron is soft.
The two Pomerons coupled to the photon vertices are hard.
\label{3pjdia}}
{The general diagram for calculating the survival probability for
diffractive production of large masses in $\gamma^*$-$p$ collisions at
high energy. The Pomerons identification is identical to the previous figure.
\label{3pspjdia}}

\DOUBLEFIGURE{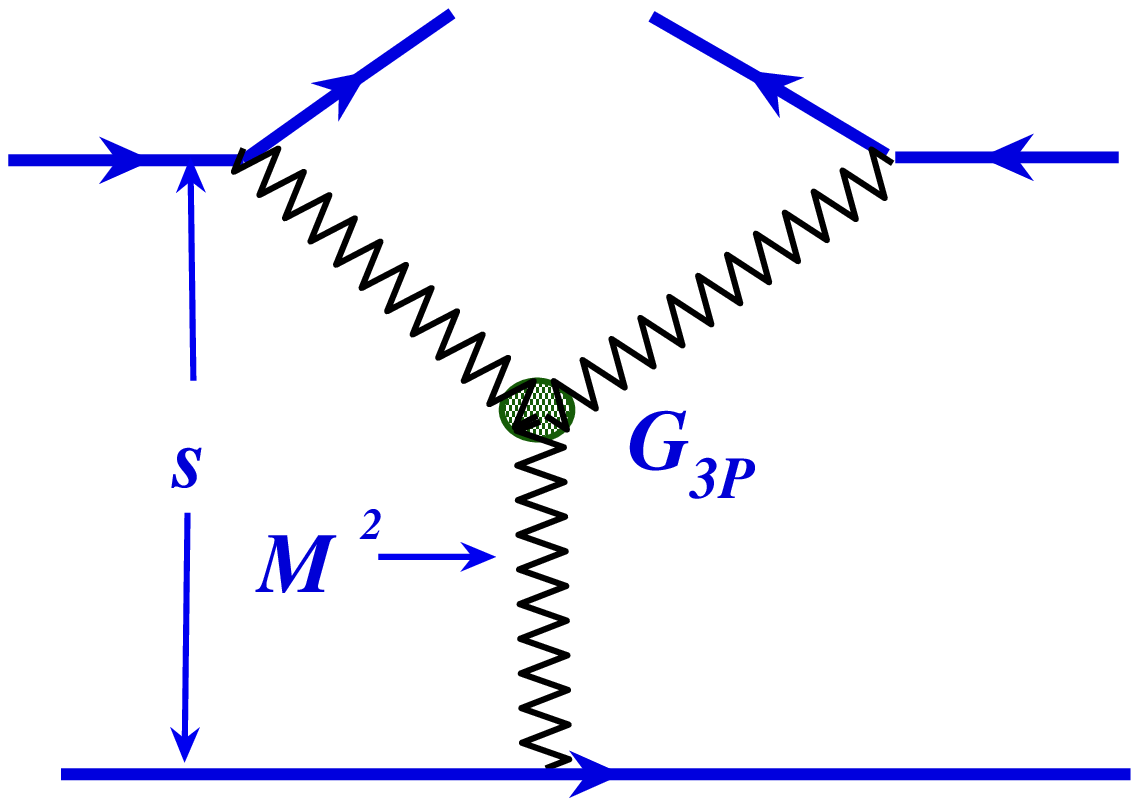,width= sp3p,width=70mm,height=50mm}
{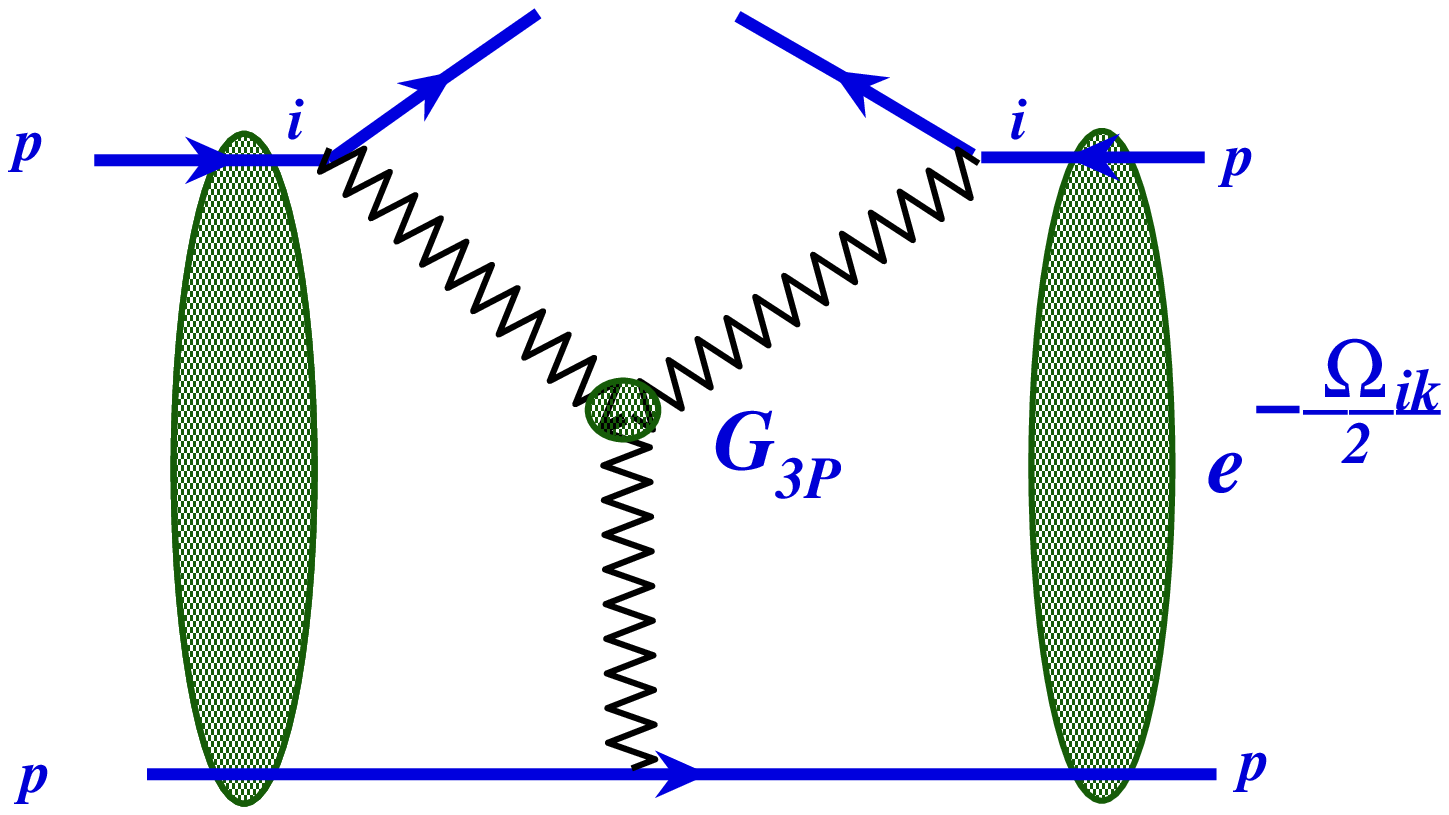,width=70mm,height=50mm}
{The general diagram for diffractive production of large masses in $p$-$p$
collisions at high energy. The zigzag lines denote soft Pomerons.
\label{3pdia}}
{The general diagram for calculating the
survival probability for diffractive production of large masses in
$p$-$p$ collisions at high energy.
The zigzag lines denote soft Pomerons.
\label{3pspdia}}
%%%%%%%%%%%%%%%%%%%%%%%%%%%%%%%%%%%%%%%%%%%%%%%%%%%%%%%%%%%%%%%%%%%%%%%%%%
The purpose of this paper is to re-examine the  procedure to extract the value of
$G_{3P}$, the triple Pomeron 'bare' coupling constant. As a by-product we assess
the stability of the survival probabilities obtained in  two and three amplitude
eikonal rescattering models.
The following topics are addressed:
\begin{enumerate}
\item We estimate the survival probability values for the triple Pomeron vertex
in $p$-$p$ high mass single diffractive (SD) collisions.
\item We investigate the difference between the survival probabilities associated
with high mass SD and the inclusive SD channels.
\item The above calculations are carried out in the two channel
GLM model\cite{2CH,3CH,heralhc} in which we compare the output of its two and
three amplitude representations,
so as to assess the reliability of our results.
\item We calculate the survival probability for the reaction given by \eq{REACT}.
Note that in an hadronic high mass SD reaction the 3P coupling consists of
three
soft Pomerons. \eq{REACT} is a hard process in which two hard Pomerons
couple to a soft one. It is possible, but not necessary, that the above 3P
couplings are equal.
\end{enumerate}

The plan of our paper is as follows:
In the next section we calculate $S^2_{spectator}(s)$ for
high  mass diffraction in $p$-$p$ scattering in two\cite{2CH} and
three\cite{3CH} amplitude models for the soft interactions.
The models are specified and their outputs compared.
In section 3 we present our estimates for the survival probability of the
process of \eq{REACT}. These calculations, carried out in a dedicated
model\cite{KORM}, show that even though the $J/\Psi$
survival probabilities are significantly larger than those calculated for
$p$-$p$,
they cannot be neglected. We note that, the reliability and accuracy of
the $J/\Psi$ calculations are considerably better than in the $p$-$p$ channel.
In the conclusions, we summarize our main results and specify some
remaining  problems.
%%%%%%%%%%%%%%%%%%%%%%%%%%%%%%%%%%%%%%%%%%%%%%%%%%%%%%%%%%%%%%%%
\section{Survival probability for the triple Pomeron vertex in proton-proton
 collisions}
%%%%%%%%%%%%%%%%%%%%%%%%%%%%%%%%%%%%%%%%%%%%%%%%%%%%%%%%%%%%%%%
\subsection{Survival probability in the eikonal model}
%%%%%%%%%%%%%%%%%%%%%%%%%%%%%%%%%%%%%%%%%%%%%%%%%%%%%%%%%%%%%%%%
The cross section for diffractive dissociation in the region of large $M$
can be viewed as a Mueller diagram (\fig{3pdia}) which can be
rewritten in terms of the triple Pomeron vertex (see Ref.\cite{3PP}).
We denote this cross section $\sigma^{3P}$ and its corresponding survival
probability, at a given $M^2$, $S^2_{3P}(M^2)$.
\beq \label{XS3P}
M^2\,\frac{d \sigma^{3P}}{d t \,d M^2}\,\,\,
=\,\,\,\frac{g^2_p(t)\,g_p(q^2=0)\,G_{3P}(t)}{16\,\pi^2}\,
\,\Lb \frac{s}{M^2}\,\Rb^{2 \alpha_P(t) - 2}\,\Lb \,\frac{M^2}{s_0}
\Rb^{\alpha_P(q^2=0) - 1},
\eeq
where $g(t)$ describe the vertex of Pomeron-proton interaction, and
$G_{3P}$ stands for the triple Pomeron vertex.
However, this diagram does not take into account the possibility of
additional rescatterings of the interacting particles
shown in \fig{3pspdia}.
The result can be written as
\beq \label{XS3P1}
M^2\,\frac{d \sigma^{3P}}{d t \,d M^2}\,
=\,S^2_{3P}(M^2)\,\,\frac{g^2_p(t)\,g_p(q^2=0)\,G_{3P}(t)}{16\,\pi^2}\,
\,\Lb \frac{s}{M^2}\,\Rb^{2 \alpha_P(t) - 2}\,\Lb
\,\frac{M^2}{s_0} \Rb^{\alpha_P(q^2=0) - 1}.
\eeq
The survival probability factor $S^2_{3P}(M^2)$
is defined {\footnote{$S^2_{3P}(M^2)$ denotes the high mass SD survival
probability, at a given $M^2$, and is identical to $S^2_{spectator}(s)$
for this specific SD reaction.} as
\beq \label{SP1}
S^2_{3P}(M^2)\,\,=\,\,\frac{\int\,d^2\,k_t \,\,M^2\,\frac{d \sigma^{3P}}
{d\,k^2_t \,d M^2}
\Lb \fig{3pspdia} \Rb}
{\int\,d^2\,k_t\,\,M^2\,\frac{d \sigma^{3P}}{d\,k^2_t\,d M^2}\Lb
\fig{3pdia}\Rb}
\,\,\,,\,\,\,\,\,\,\,\,\,\mbox{with}\,\,\,\,t\,=\,\,- \,k^2_t.
\eeq

The easiest way to calculate the diagram of \fig{3pspdia} is to
first transform
 the diagram of \fig{3pdia} to impact parameter space.
This is done by introducing the momentum $q$ along the lowest Pomeron in
\fig{3pdia}. In this case,
\beq \label{SP2}
T\Lb s,M^2; q \Rb\,\,\equiv\,\,\int\,d^2\,k_t\,\,M^2\,
\frac{d \sigma^{3P}}{dk^2_t\,d M^2} \Lb \fig{3pdia} \Rb
\,\,\,\longrightarrow\,
\eeq
$$
\int\,d^2k_t\,\,\frac{g_p(k^2_t)\,g_p((\vec{k} - \vec{q})^2_t)\,g_p(q^2_t)\,
G_{3P}(k^2_t,(\vec{k} - \vec{q})^2_t,q^2 )}{16\,\pi^2}\,
\,\Lb \frac{s}{M^2}\,\Rb^{\alpha_P(k^2_t)\,+\,\alpha_P((\vec{k} - \vec{q})^2_t)\,-\,2}
\,\Lb \,\frac{M^2}{s_0} \Rb^{\alpha_P(q^2) - 1}\,.
$$
From \eq{SP2} we find the form of this amplitude
in impact parameter space to be
\beq \label{SP3}
T\Lb s,M^2; b \Rb\,\,\equiv\,\,\int\,\frac{d^2 q}{(2 \pi)^2}\,\,A\Lb s,M^2;q\Rb.
\eeq
Using a linear approximation for the Pomeron trajectory and a Gaussian
form for all vertices
\beq \label{SP4}
\alpha_P(t)=1\,+\,\Delta\,\alpha'_P\,t,\,
\,\,\,\,\,g_p(k^2)=g_p(0)\,e^{- b_p\,k^2},\,
\,\,\,\,\,G_{3P}(k_1,k_2,k_3)=G_{3P}(0,0,0)\,
e^{ - b_p (k^2_1\,+\,k^2_2\,+\,k^2_3)}\,,
\eeq
we obtain
\beq \label{SP5}
T\Lb s,M^2; b \Rb\,=\,\frac{g_{3P}}{16\pi^2}\,
\nu(\xi)\,\nu(\xi)\,\nu(y)\,\frac{\pi}{d(\xi)+d(\xi)+d(y)}\,
\exp\Lb - \frac{d(y)\,[d(\xi)+d(\xi)]}{d(\xi)+d(\xi)+d(y)}\,b^2 \Rb.
\eeq
\bea\label{SP61}
y\,=\,\ln \Lb M^2/s_0 \Rb,\,\,\,\,\,\,\xi\,=\,\ln \Lb s/M^2 \Rb,
\,\,\,\,\,\,g_{3P}\,\,\equiv\,\,G_{3P}(0,0,0)/g_p(0),\,\,\,\,\,\,
\nu(y)\,=\,\frac{g^2_p(0)}{\pi\bar{R}^2(y)}\,e^{\Delta\,y},
\eea
where
\beq\label{SP62}
\bar{R}^2(y)\,\,=\,\,2\,R^2_0\,\,+\,\,2\,r^2_0\,+\,4\,\alpha'_P\,y
\,\,\,\,\,\,\mbox{and}\,\,\,\,\,\,d(y)\,\equiv\,\frac{1}{\bar{R}^2(y)}.
\eeq
$r^2_0$ denotes the radius of the triple Pomeron vertex.
In the following we take \cite{1CH} $r^2_0\,=\,0.5\,GeV^{-2}$.

Using \eq{SP5}, the expression for the survival probability
(see \eq{SP1}) in a simple eikonal model,
accounting for the rescattering corrections, can be written as
\beq \label{SP7}
S^2_{3P}(M^2)\,=\,\frac{\int\,d^2 b\,T\Lb s,M^2;b\Rb\,
\exp\Lb\,-\,\Omega\Lb\xi+y;b\Rb\,\Rb}{\int\,d^2 b\,T\Lb s,M^2;b \Rb},\,\,\,\,\,
\mbox{where}\,\,\,\,\,\Omega\Lb \xi+y;b \Rb\,=\,\nu_{pp}\,
e^{-\frac{b^2}{R_{pp}(\xi+y)}}.
\eeq
\beq \label{SP8}
\nu_{pp}\,=\,\frac{g^2_p(0)}{\pi  R_{pp}(\xi+y)}\,
e^{\Delta (\xi+y)}\,\,\,\,\,\mbox{and}\,\,\,\,\,
R_{pp}(\xi+y)\,=\,4\,R^2_{0,p}\,+\,4\,\alpha'_P\,(\xi y).
\eeq
%%%%%%%%%%%%%%%%%%%%%%%%%%%%%%%%%%%%%%%%%%%%%%%%%%%%%%%%%%%%
\subsection{Two channel models: main ideas and formulae }
%%%%%%%%%%%%%%%%%%%%%%%%%%%%%%%%%%%%%%%%%%%%%%%%%%%%%%%%%%%%
In the eikonal model only elastic rescatterings have been taken into account.
Two channel eikonal
models have been developed so as to also include  rescatterings through
diffractive dissociation (see Refs.\cite{2CH,3CH,heralhc} and references
therein).
In this formalism, diffractively produced hadrons at a given vertex are
considered as a single hadronic state
described by the wave function $\Psi_D$, which is orthonormal
to the wave function $\Psi_h$ of the hadron (proton in the case of interest),
$<\Psi_h|\Psi_D>\,=\,0 $.

Introducing two wave functions that diagonalize the  2x2
interaction matrix
${\bf T}$
\beq \label{SP9}
A_{i,k}\,=\,<\Psi_i\,\Psi_k|\mathbf{T}|\Psi_{i'}\,\Psi_{k'}>\,\,=
\,\,A_{i,k}\,\delta_{i,i'}\,\delta_{k,k'},
\eeq
In our past publications we referred to the GLM eikonal models
according to the number of the rescattering channels considered, i.e.
elastic\cite{1CH},
elastic + SD\cite{2CH} and elastic + SD + DD\cite{3CH}. In retrospect, we
consider it more appropriate to define these models according to the
dimensionality of their base. We, therefore, call the above a two
channel model, making the distinction between its two and three amplitude
representations.

We can rewrite the amplitude $A_{i,k}$ in a form that satisfies
the unitarity constraints
\beq \label{SP10}
A_{i,k}(s,b)\,=\,i\,\Lb \,1 \,-\,\exp\Lb - \frac{\Omega_{i,k}(s,b)}{2}\Rb\Rb.
\eeq
In this formalism we have
\beq \label{SP10a}
G^{in}_{i,k}(s,b)\,=\,1\,-\,\exp\Lb - \Omega_{i,k}(s,b)\Rb.
\eeq
$G^{in}$ is the probability for all inelastic interactions in the
scattering of particle $i$ off particle $k$.
From \eq{SP10a} we deduce that the probability that the initial projectiles
reach the interaction unchanged, regardless of the initial state rescatterings,
is $\exp \Lb - \Omega_{i,k}(s,b) \Rb$.

In this representation the observed states can be written in the form
\beq \label{SP11}
\Psi_h\,=\,\alpha\,\Psi_1\,+\,\beta\,\Psi_2\,,
\,\,\,\,\,\,\,\,\Psi_D\,=\,-\,\beta\,\Psi_1\,+\,\alpha \,\Psi_2\,,
\,\,\,\,\,\,\,\,\mbox{where}\,\,\,\,\,\,\,\alpha^2\,+\,\beta^2\,\,=\,\,1.
\eeq
The obvious generalization of \eq{SP5} is
\beq \label{SP12}
T\Lb s,M^2;b \Rb\,=\,\sum_{i,k,l}\,<p|l>^2\,<p|k>\,
T^{l,i}_{k}\Lb s,M^2;b \Rb\,<p|k>\,<p|i>^2,
\eeq
where $<p|1>\,=\,\alpha$ and $<p|2>\,=\,\beta$.
\beq \label{SP13}
T^{l,i}_k\Lb s,M^2;b \Rb\,=\,\frac{g_{3P}}{16\pi^2}\,
\nu_l(\xi)\,\nu_i(\xi)\,\nu_k(y)\,\frac{\pi}{d_l(\xi)+d_i(\xi)+d_k(y)}\,
\exp\Lb - \frac{d_k(y)\,[d_l(\xi)+d_i(\xi)]}{d_l(\xi)+d_i(\xi)+d_k(y)}\,b^2 \Rb,
\eeq
where $g_{3P}\,=\,G_{3P}/g_1(0)$ and
\bea
\nu_k(y)\,=\,\frac{g_k(0)g_1(0)}{\pi\bar{R}^2_k(y)}\,e^{\Delta \,y}\,\,
&\mbox{and}&\,\bar{R}^2_k(y)\,=\,2\,R^2_{0,k}\,+\,2\,r^2_0\,+\,4\,\alpha'_P\,y,
\,\,\,\,\,\,d_k(y)\,\equiv\,\frac{1}{\bar{R}^2_k(y)}\,.
\label{SP131}
\eea
The numerator of \eq{SP7} is written in this representation as
\beq \label{SP14}
\int d^2\,k_t\,M^2\,\frac{d\sigma^{3P}}{dk^2_t\,d M^2}\Lb
\fig{3pspdia}\Rb\,=
\eeq
$$
\,\,\int\,d^2b\,\sum_{i,k,l}\,<p|l>^2\,<p|k>\,e^{-\frac{\Omega_{l,k}(s,b)}{2}}\,\,
T^{l,i}_{k}\Lb s,M^2; b \Rb\,
e^{-\frac{\Omega_{i,k}(s,b)}{2}}\,<p|k>\,<p|i>^2.
$$
For $\Omega_{i,k}(s,b)$ we take
\beq \label{SP15}
\Omega_{i,k}(s,b)\,=\,\nu_{i,k}\,e^{-\frac{b^2}{R^2_{i,k}(\xi+y)}}\,,
\eeq
where
\beq \label{SP16}
\nu_{i,k}\,=\,\frac{g_i(0)\,g_k(0)}{\pi\,R^2_{i,k}(\xi+y)}\Lb
\frac{s}{s_{0}}\Rb^{\Delta},\,\,\,\,\
\mbox{and}\,\,\,\,\,R^2_{i,k}(\xi+y)\,=\,2\,R^2_{0,i}+2\,R^2_{0,k}\,
+\,4\,\alpha'_P\,(\xi+y)\,.
\eeq
In the following we denote $g^2_{i,k}\,=\,g_i(0)g_k(0)$.

In our calculations we take \cite{2CH} $R^2_{0,1}$ as a free parameter
while $R^2_{0,2} = 0$.
The survival probability can be calculated as the ratio
\beq \label{SP17}
S^2_{3P}(M^2)\,=\,\frac{\int\,d^2\,b\,\,N(\xi,y;b)}{\int\,d^2\,b\,\,D(\xi,y; b)},
\eeq
where
\bea \label{SPN}
N(\xi,y;b)\,&=&\,\alpha^6\,T^{1,1}_1(b)\,e^{-\Omega_{1,1}(b)}\,+
\,2\,\alpha^4\,\beta^2\,T^{1,2}_1(b)\,e^{-\frac{\Omega_{1,1}(b)+\Omega_{1,2}(b)}{2}}
\,+\,\alpha^2\,\beta^4\,T^{2,2}_1(b)\,e^{-\Omega_{1,2}(b)}\nonumber \\
& &+\,\alpha^4\,\beta^2\,T^{1,1}_2(b)\,e^{-\Omega_{1,2}(b)}\,+
\,2\,\alpha^2\,\beta^4\,T^{1,2}_2(b)\,e^{-\frac{\Omega_{2,2}(b)
+\Omega_{1,2}(b)}{2}}\,+\,\beta^6\,T^{2,2}_2(b)\,e^{-\Omega_{2,2}(b)}\,,
\eea
and
\bea \label{SPD}
D(\xi,y;b)\,&=&\,\alpha^6\,T^{1,1}_1(b)\,+
\,2\,\alpha^4\,\beta^2\,T^{1,2}_1(b)\,+
\,\alpha^2\,\beta^4\,T^{2,2}_1(b)\,\,\nonumber\\
& & \,+ \,\alpha^4\,\beta^2\,T^{1,1}_2(b)\,+
\,2\,\alpha^2\,\beta^4\,T^{1,2}_2(b)\,+
\,\beta^6\,T^{2,2}_2(b)\,.
\eea

For completeness we also present  the integrated cross sections
of the diffractive channels in the two channel model, together
with the corresponding elastic and total cross sections.
The amplitudes for the elastic and the diffractive channels
have the following form \cite{2CH,3CH}
\bea
a_{el}\Lb s,b\Rb\,&=&\,i\,\Lb
\,1\,-\alpha^4\,e^{-\frac{\Omega_{1,1}(s,b)}{2}}\,-
\,2\,\alpha^2\,\beta^2\,e^{-\frac{\Omega_{1,2}(s,b)}{2}}\,-
\,\beta^4\,e^{-\frac{\Omega_{2,2}(s,b)}{2}}\Rb, \label{EL}\\
a_{sd}\Lb s,b\Rb\,&=&\,i\,\alpha\,\beta
\Lb\,\,\alpha^2\,e^{-\frac{\Omega_{1,1}(s,b)}{2}}\,-
\,(\alpha^2 \,-\,\beta^2)\,
e^{-\frac{\Omega_{1,2}(s,b)}{2}}\,-\,
\,\beta^2\,e^{-\frac{\Omega_{2,2}(s,b)}{2}}\,\Rb, \label{SD}\\
a_{dd}\Lb s,b\Rb \,&=&\,i\,\alpha^2\,\beta^2
\Lb\,-\,e^{-\frac{\Omega_{1,1}(s,b)}{2}}\,+
\,2\,e^{-\frac{\Omega_{1,2}(s,b)}{2}}\,-
\,e^{-\frac{\Omega_{2,2}(s,b)}{2}}\,\Rb. \label{DD}
\eea

Using \eq{SP15} and \eq{SP16}, as well as the general
cross section formulae we get
\beq \label{XS1}
\sigma_{tot}(s)\,=\,2\int\,d^2 b \,a_{el}\Lb s,b\Rb,
\,\,\,\,\,\,\sigma_{el}(s)\,=\,\int\,d^2 \,b \,\,|a_{el}\Lb s,b\Rb|^2,
\eeq
%$$
\beq \label{XS2}
\sigma_{sd}(s)\,=\,\int\,d^2 \,b \,|a_{sd}\Lb s,b\Rb|^2,
\,\,\,\,\,\,\sigma_{dd}(s)\,=\,\int\,d^2 \,b \,|a_{dd}\Lb s,b\Rb|^2.
\eeq

It is instructive to present the calculation for the diffractive channels in
the form of a survival probability,
which we define as the ratio of the output corrected diffractive cross section
to the input non corrected cross section.
\bea \label{SPSD}
S^2_{sd}\,=
\,\frac{\int\,d^2\,b\,|a_{sd}(s,b)|^2}{\int\,d^2\,b\,|a^1_{sd}(s,b)|^2}\,,
\eea
where
\bea \label{SPSDa}
a^1_{sd}\Lb s,b\Rb \,=\,\frac{i\,\alpha\,\beta}{2}
\Lb-\,\alpha^2\,\Omega_{1,1}(s,b)\,+\,(\alpha^2 \,-\,\beta^2)\,
\Omega_{1,2}(s,b)\,+\,\beta^2\,\Omega_{2,2}(s,b)\Rb.
\eea
We will discuss the results and interpretation
of our calculations in the next subsection.
%%%%%%%%%%%%%%%%%%%%%%%%%%%%%%%%%%%%%%%%%%%%%%%%%%%%%%%%%%%%%%%%
\subsection{Three models used for fitting the experimental data}
%%%%%%%%%%%%%%%%%%%%%%%%%%%%%%%%%%%%%%%%%%%%%%%%%%%%%%%%%%%%%%%%
 To calculate the survival probabilities one needs to
specify  the
opacities $\Omega_{i,k}(s,b)$. These are determined from a global fit of
the
experimental soft scattering data.
We have used three models based on
the general formulae given in \eq{SP13} - \eq{SP16}, but with different
input assumptions.
Note that the above global fit has in addition to the
Pomeron contribution, also a secondary Regge sector (see Ref.\cite{2CH}).
This is necessary as the data base contains many experimental points
from lower ISR energies.
 A  study of the Pomeron
component alone, without a Regge contribution, is not
possible at this time, since the corresponding high energy sector of the
data
base is too small to constrain the fitted parameters.
The Regge parameters are not
quoted in this paper and will be discussed in detail in a forthcoming
publication.
%%%%%%%%%%%%%%%%%%%%%%%%%%%%%%%%%%%%%%%%%%%%%%%%%%%%%%%%%%%%%%%%%
\subsubsection{Two amplitude model}
%%%%%%%%%%%%%%%%%%%%%%%%%%%%%%%%%%%%%%%%%%%%%%%%%%%%%%%%%%%%%%%%%
In this  model, denoted by  Model A \cite{2CH} we assume that the
double diffraction cross section
is negligible, and we take $a_{dd}$ in \eq{DD} to be zero.
This allows us to express $\Omega_{2,2}$
in terms of $\Omega_{1,1}$ and
$\Omega_{1,2}$, leading to the following formulae (see Refs.\cite{2CH,3CH,heralhc}):
\bea
a_{el}(s,b )\,=\,i \Lb 1 - \exp\Lb - \frac{\Omega_{1,1}(s,b)}{2}\Rb-
\,2\,\beta^2\,\exp\Lb - \frac{\Omega_{1,1}(s,b)}{2}\Rb\,
\Lb 1\,-\,\exp\Lb - \frac{\Delta \Omega(s,b)}{2}\Rb \Rb \Rb,  \label{A1}
\eea
\bea
a_{sd}(s,b )\,=\,- i\,\alpha\,\beta \,\exp\Lb - \frac{\Delta \Omega(s,b)}{2}\Rb\,
\Lb 1\,-\,\exp\Lb - \frac{\Delta \Omega(s,b)}{2}\Rb\Rb \,.\label{A2}
\eea
The above is a two amplitude model  with two opacities $\Omega_{1,1}$
and $\Delta \Omega\,=\,\Omega_{1,2}\,-\,\Omega_{1,1}\,. \label{A3}$
Following Ref.\cite{2CH},
we assume both $\Omega_{1,1}$ and $\Delta \Omega$ to be Gaussian in b.
\bea
\Omega_{1,1}(s,b)\,=\,\frac{g^2_{1,1}}{\pi R^2_{1,1}(s)}\,\Lb
\frac{s}{s_0}\Rb^{\Delta}\,\exp\Lb -\,\frac{b^2}{R^2_{1,1}(s)}\Rb\,,\label{A4}
\eea
\bea
\Delta \Omega(s,b)\,=\,\frac{g^2_{\Delta}}{\pi R^2_{\Delta}(s)}
\,\Lb
\frac{s}{s_0}\Rb^{\Delta}\,\exp\Lb -\,\frac{b^2}
{R^2_{\Delta}(s)}\Rb\,.\label{A5}
\eea
Note that in this two amplitude model $R^2_{\Delta}$ is the radius of $\Delta
\Omega(s,b)$. As we shall see, in the three amplitude model
$R^2_{1,2}$ is the radius of $\Omega_{1,2}$.
The radii $R^2_{1,1}$ and $R^2_{1,2}$ are specified in \eq{SP16}.
We have also studied a two amplitude model in which both $\Omega_{1,1}$ and
$\Omega_{1,2}$ are Gaussian in $b$. The output obtained in this two amplitude
model is compatible with the output of Model A. This is a consequence of the fit,
discussed below, in which consistently $\Omega_{1,2}\,\gg\,\Omega_{1,1}$.
%%%%%%%%%%%%%%%%%%%%%%%%%%%%%%%%%%%%%%%%%%%%%%%%%%%%%%%%%%%%%%%%%%%%%%
\subsubsection{Three amplitudes models}
%%%%%%%%%%%%%%%%%%%%%%%%%%%%%%%%%%%%%%%%%%%%%%%%%%%%%%%%%%%%%%%%%%%%%%
In the  three amplitude models  we do not make any assumptions
regarding the value of the double diffraction cross sections\cite{DDD}
which are contained in our  data base.
We use \eq{SP15} and \eq{SP16} to parameterize the three independent
opacities:
$\Omega_{1,1}$, $\Omega_{1,2}$ and $\Omega_{2,2}$,
which are all taken to be Gaussian in $b$.
For details see Ref.\cite{3CH}.
Assuming Regge factorization, $\Omega_{1,2}$ is determined by
$\Omega_{1,1}$ and $\Omega_{2,2}$, see \eq{SP16}).
We denote this model B(1). As we shall see, Model B(1) does not reproduce
the data well.
We have, thus, also examined a non factorizable model, denoted B(2), in which
 $\nu_{1,2}$ of \eq{SP16} has been replaced
by the expression
\beq \label{SP16C}
\nu_{1,2}\,\,=\,\,\frac{g^2_{1,2}}{\pi R^2_{1,2}(\xi + y)}\,
\Lb \frac{s}{s_0}\Rb^{\Delta}\,,
\eeq
where $g^2_{1,2}$ is a free parameter. This additional degree of freedom
violates  Regge factorization for the input Pomeron,
but it allows us to describe the experimental data on the
double diffraction cross section which we failed to fit in model B(1).

%%%%%%%%%%%%%%%%%%%%%%%%%%%%%%%%%%%%%%%%%%%%%%%%%%%%%%%%%%%%%%%%%%%%%%
\subsection{Results}
%%%%%%%%%%%%%%%%%%%%%%%%%%%%%%%%%%%%%%%%%%%%%%%%%%%%%%%%%%%%%%%%%%%%%
The parameters of Model A , quoted from Ref.\cite{2CH} are based on
a fit to  55 experimental data points base
which include the $p$-$p$ and $\bar{p}$-$p$ total cross sections,
integrated elastic cross sections,
integrated single diffraction cross sections,
and the forward slope of the elastic cross section
in the ISR-Tevatron energy range.
As stated, we  neglected the (very few) reported DD cross sections.
The fitted parameters of Model A are listed in Table 1
with a corresponding $\chi^2/(d.o.f)$ of 1.50.

The fit to Models B(1) and B(2) are   based on
the formulae given in the previous subsection and in Refs. \cite{3CH,heralhc}.
The new data base includes the data used for Model A,
plus 5 double diffraction cross sections points \cite{DDD}.
The values of the parameters and $\chi^2$ associated with
B models are also given in Table 1.
The best fit obtained from Model B(1) has a $\chi^2/(d.o.f) = 2.3$,
which is unsatisfactory. This is manifested not only by the
high $\chi^2$, but also by inability of the model to reproduce the
$\sigma_{dd}$  points, which are severely underestimated.
We thus consider Model B(2) in which the couplings of the
three amplitudes are independent. This model produces a much better
fit with  $\chi^2/(d.o.f) = 1.25$, as well as reproducing the
experimental results for $\sigma_{dd}$. The cross section
predictions of Model B(2) are shown in \fig{spsd}.
The parameters associated with Models A, B(1) and B(2) were
used to calculate the survival probability for two different soft SD final states.
$S^2_{3P}$ corresponds to single diffraction dissociation in the high mass region
(see \eq{SP17}). We note that $S^2_{3P}$  is weakly dependent on large
$M^2$ - mass of produced hadrons. $S^2_{sd}$ is the survival probability
corresponding to the entire region of the produced diffractive mass (see \eq{SPSD}).
The calculated survival probabilities are presented in \fig{sppp}.
We note a significant difference between the various outputs which will
be discussed below.

The following are a few qualitative characteristic remarks and comments:
\begin{enumerate}
\item \quad Clearly, our analysis favors the non factorizable input of
Models A and B(2) over the factorizable input of Model
B(1), which is theoretically more appealing. Technically,
the factorization breaking is induced by the output results,
in which $g_2$ is considerably larger than $g_1$ resulting
in $A_{1,2} \,\gg\,A_{1,1}$. As we have shown in Ref.\cite{3CH},
this is a consequence of the striking experimental observation that in
$p$-$p$ (and $\bar p$-$p$) scattering
$R_d \,=\,(\sigma_{el} + \sigma_{sd} + \sigma_{dd})/\sigma_{tot}\,\approx \,0.4$.
This is different from $R_{el} = \sigma_{el}/\sigma_{tot}$ which
rises monotonically with energy. The analysis made in Ref. \cite{3CH}
found that to be consistent with the experimental behaviour of $R_D$
and $R_{el}$ requires that $g_2 > 10\,g_1$. The best value
obtained is $g_2 = \sqrt{300}g_1$ which is very close to the values
found in Model B(2).

\item \quad  We observe a general systematic behaviour in which the 3P
and SD survival probabilities become smaller when we add an amplitude to
the initial state rescattering chain. In the context of this paper
we find that $S^2(\mbox{Model A})\,\gg\,S^2(\mbox{Model B})$
for both 3P and SD channels.

\FIGURE[ht]{
\centerline{\epsfig{file= 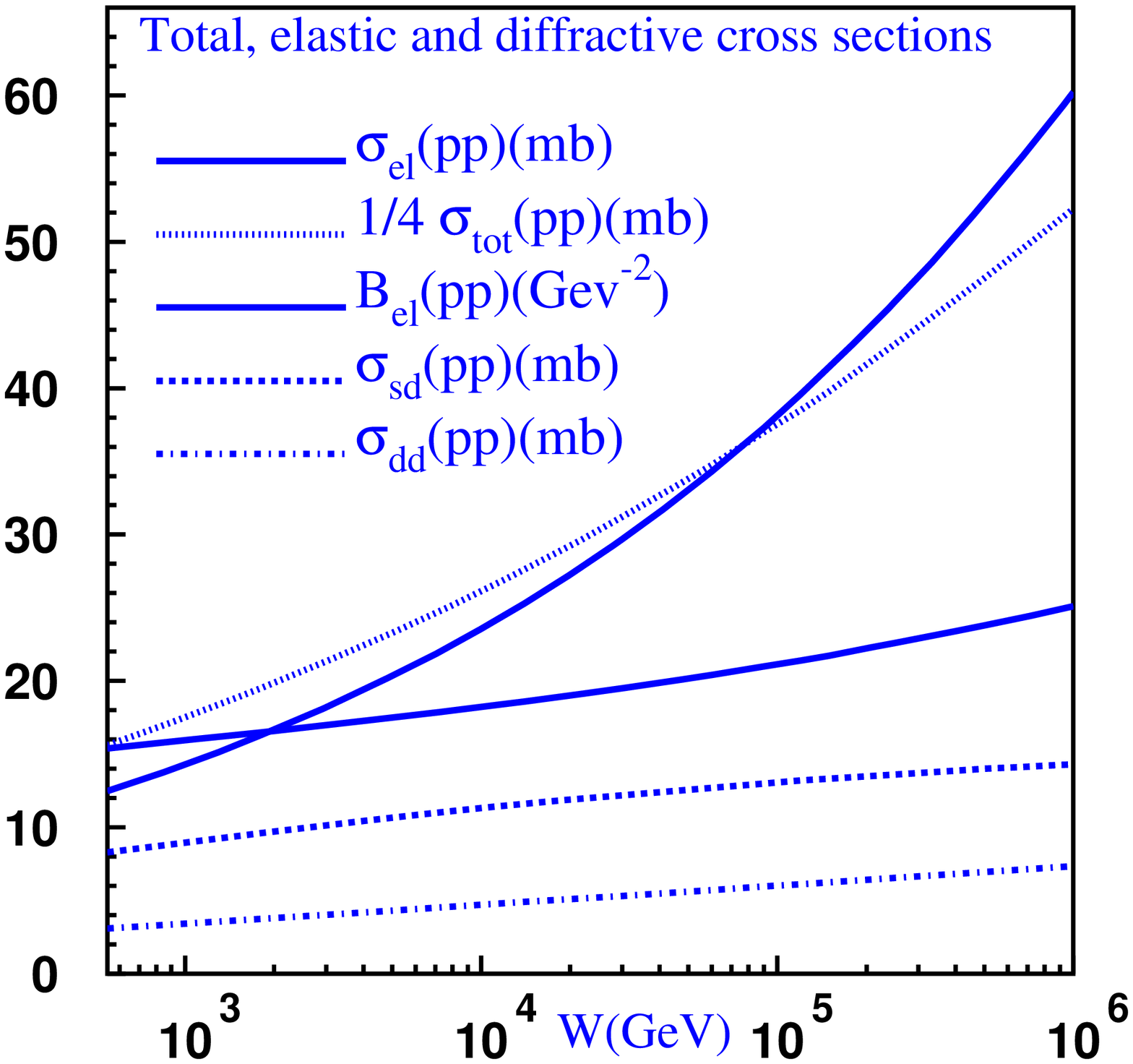,width=140mm,height=110mm}}
\caption{Total, elastic and diffractive dissociation cross sections in model
B(2).}
\label{spsd}}
%%%%%%%%%%%%%%%%%%%%%%%%%%%%%%%%%%%%%%%%%%%%%%%%%%%%%%%%%%%%%%%%%%%%%

%%%%%%%%%%%%%%%%%%%%%%%%%%%%%%%%%%%%%%%%%%%%%%%%%%%%%%%%%%%%%%%%%%%%%%%%%%
\begin{footnotesize}
\TABLE[ht]{
\begin{tabular}{|l|l|l|l|l|l|l|l|l|c|}
\hline
Model & $ \Delta $ & $\beta$ & $R^{2}_{0;1,1}$ & $\alpha^{\prime}_{P}$&
$g^2_{1,1}$ & $g^2_{2,2}$ & $g^2_{\Delta}$ & $g^2_{1,2}$ \\
\hline
A & 0.126 & 0.464 &16.34 $GeV^{-2}$ &0.200 $GeV^{-2}$ &12.99 $GeV^{-2}$&
N/A & 145.6 $GeV^{-2}$& N/A \\
\hline
B(1)& 0.150 & 0.526 &20.80 $GeV^{-2}$& 0.184 $GeV^{-2}$ &4.84 $GeV^{-2}$&
4006.9 $GeV^{-2}$& N/A &139.3 $GeV^{-2}$\\
\hline
B(2) & 0.150 & 0.776 &20.83 $GeV^{-2}$& 0.173 $GeV^{-2}$ &9.22 $GeV^{-2}$&
3503.5 $GeV^{-2}$& N/A &6.5 $GeV^{-2}$ \\
\hline
\end{tabular}
\caption{Fitted parameters for Models A, B(1) and B(2).
$R^{2}_{0;1,2}=R^{2}_{0;\Delta}=0.5 R^{2}_{0;1,1}$, $R^{2}_{0;2,2}=0.$ }}
\end{footnotesize}
%%%%%%%%%%%%%%%%%%%%%%%%%%%%%%%%%%%%%%%%%%%%%%%%%%%%%%%%%%%%%%%%%%%%%%%%%

%%%%%%%%%%%%%%%%%%%%%%%%%%%%%%%%%%%%%%%%%%%%%%%%%%%%%%%%%%%%%%%%%%
\FIGURE[ht]{
\centerline{\epsfig{file= 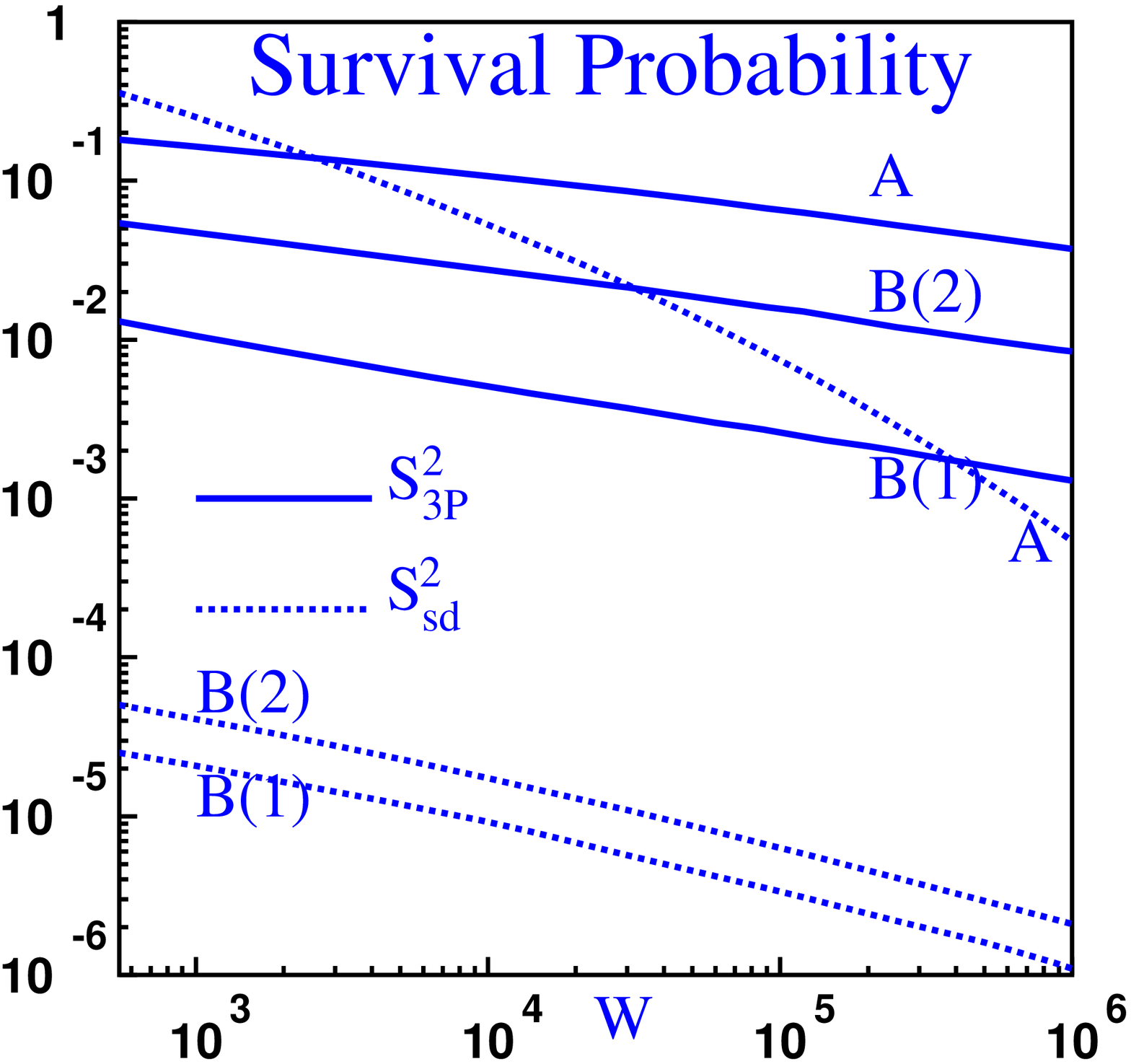,width=110mm,height=90mm}}
\caption{Survival probabilities for single diffraction $p$-$p$
collision. $S^2_{3P}$ (solid line) denotes the survival probability for
high mass diffraction dissociation (see \eq{SP17}).
$S^2_{sd}$ (dotted line) is the survival probability for
diffractive dissociation in the entire kinematic region (see \eq{SPSD}.
The upper curves (solid and dotted) refer to  Model A\cite{2CH}.
The lower curves relate to Model B of Ref.\cite{3CH}.}
\label{sppp}}
%%%%%%%%%%%%%%%%%%%%%%%%%%%%%%%%%%%%%%%%%%%%%%%%%%%%%%%%%%%%%%%%%%%%%

 We trace this dramatic difference to
the relaxation of the constraints  on the initial
rescatterings. Allowing an
additional amplitude, enables  increased screening of the input amplitude.
\item  \quad An important observation, correlated to the above, is that the
unscreened input cross sections
of Model B(2) are  considerably larger than the unscreened input
cross sections of Model A. This is a consequence of the large difference
between the fitted values of
$g^2_{2,2}$ of Model B(2) and $g^2_{\Delta}$ of Model A, and, also, the
different values of $\Delta$ (the Pomeron intercept) of these models.
Since these  models  provide
compatible good reproductions of the fitted data base, we conclude that the
large difference in the unscreened cross sections of Models A and B(2) are
compensated by the reciprocal difference in the corresponding survival
probabilities.
\item \quad The survival probabilities calculated in our two amplitude Model A are
in agreement with those calculated by Khoze, Martin and Ryskin in
their two channel model\cite{DG3P}, which differs from ours.
Our present observation that the three amplitude Model B
results in considerably smaller
$S^2$ values, implies that the presumed consistency between most of the published
survival probability outputs (for details see Ref.\cite{heralhc}) should be
carefully re-examined using more robust models.
\item \quad $S^2_{3P}$ are consistently higher than $S^2_{sd}$. Note that the input of
$\sigma_{sd}(\mbox{high mass})$, described by the triple Pomeron diagram,
behaves as $s^{\Delta}$. On the other hand $\sigma_{sd}(\mbox{low  mass})$,
for which the triple Pomeron diagram is not applicable, behaves as  $s^{2\,\Delta}$.
\end{enumerate}

Diffractive processes are very important at cosmic ray energies.
Kama et al.\cite{KAK} have shown that diffractive $p$-$p$ interactions
play a crucial role in understanding the spectrum of galactic gamma-rays
that come predominantly from $\pi^{0} \rightarrow \gamma \gamma$. The
inclusion of diffractive processes makes the gamma ray spectrum harder, and
when this is included together with the assumption of Feynman scaling
violations, one can explain about half of the "GeV Excess".
The magnitude of the diffractive cross sections is a crucial element of
this and similar studies.
A review of our results at cosmic ray energies will be published elsewhere.

%%%%%%%%%%%%%%%%%%%%%%%%%%%%%%%%%%%%%%%%%%%%%%%%%%%%%%%%%%%%%%%%%%%%%%%%%
\section{Survival probability for triple Pomeron vertex in J/$\Psi$-$p$
photo and DIS production}
%%%%%%%%%%%%%%%%%%%%%%%%%%%%%%%%%%%%%%%%%%%%%%%%%%%%%%%%%%%%%%%%%%%%%%%%%%
In this section we calculate the survival probabilities for high mass diffraction
in the reaction of \eq{REACT}. From \fig{3pspjdia} one can see that we need the
following ingredients to make these estimates: i) the amplitude for the
interaction of a colorless dipole with the target.; and ii) the description of
J/$\Psi$ production with no initial state interactions with the
target, shown in \fig{3pjdia}.

For the scattering dipole amplitude we take a model developed by one of
us\cite{KORM}. This model is based on the solution of a generating
functional\cite{MUCD,L1,L2}, with an additional assumption
that the dipoles do
not change their sizes during the interaction. The amplitude is equal to
\beq \label{SPJ1}
N(Y=\ln(1/x);r,b)\,=\,\frac{\Omega\Lb Y;r,b\Rb}{1\,+\,\Omega\Lb Y;r,b\Rb}\,,
\eeq
where
\bea
\Omega\Lb Y;r,b\Rb\,=\,
\frac{\pi^{2}}{N_{c}}r^2\,xG(x_{0},\mu^{2})\;
xG\Lb \frac{x}{x_{0}},\mu^2\,=\,\frac{C}{r^2}\,+\,\mu^2_0\Rb\,S(b). \label{SPJ2}
\eea
$S(b)$ is the proton b-profile,
\bea
S(b)\,=\,\frac{2}{\pi R^2}\frac{\sqrt{8}b}{R}K_1\left(\frac{\sqrt{8}b}{R}\right).
\label{SPJ3}
\eea
The saturation scale $Q^{2}_{s}(Y)$ is defined from the condition
\bea
\Omega\Lb Y; r_{s}, b\Rb\;=\;1\label{SPJ31}\;\;\;
\mbox{with}\;\; r^{2}_{s}\;=\;4/Q^{2}_{s}.
\eea

The gluon structure function $xG(x,\mu^2)$ satisfies the DGLAP
evolution equation with the initial condition
$xG(x,Q^2_0)\,=\,A/x^{\omega_0}$ at $Q^2_0\;=\;1\;GeV^2$. In other
words, \eq{SPJ2} describes the contribution of the hard Pomeron that
can be calculated in pQCD. All parameters in \eq{SPJ2} have been
found by fitting to the data on $F_2$. The fit is good and has
$\chi^2/d.o.f.\,=\,1.07$\cite{KORM}. The values of the fitted
parameters, are compatible with the parameter values obtained in
other competing models \cite{MODELS}. \eq{SPJ3} is the Fourier
transform of the electro-magnetic form factor of the proton in
t-space. \eq{SPJ1} is quite different from the eikonal approximation
that has been used in other models, and has a form which is typical
for the `fan' diagrams which are summed in the mean field
approximation (MFA).

One can see directly from
\eq{SPJ1} and \eq{SPJ2} that
\beq \label{SPJ4}
N(Y=\ln(1/x);r,b)\,\,\,\,\xrightarrow{r\,\rightarrow\,\infty;\,\;x \,\,fixed}
\,\,\,\,1\,,\,\,\,\,\,\,\,\,\,\,\,\,\,
N(Y=\ln(1/x);r,b)\,\,\,\,\xrightarrow{x\,\rightarrow\,0;
\,\,r\,\,fixed}\,\,\,\,1.
\eeq
In \fig{F2} we show how \eq{SPJ1} fits the experimental data.
%%%%%%%%%%%%%%%%%%%%%%%%%%%%%%%%%%%%%%%%%%%%%%%%%%%%%%%%%%%%%%%%%%%%%
\FIGURE[ht]{
\begin{tabular}{c c c}
\epsfig{file= 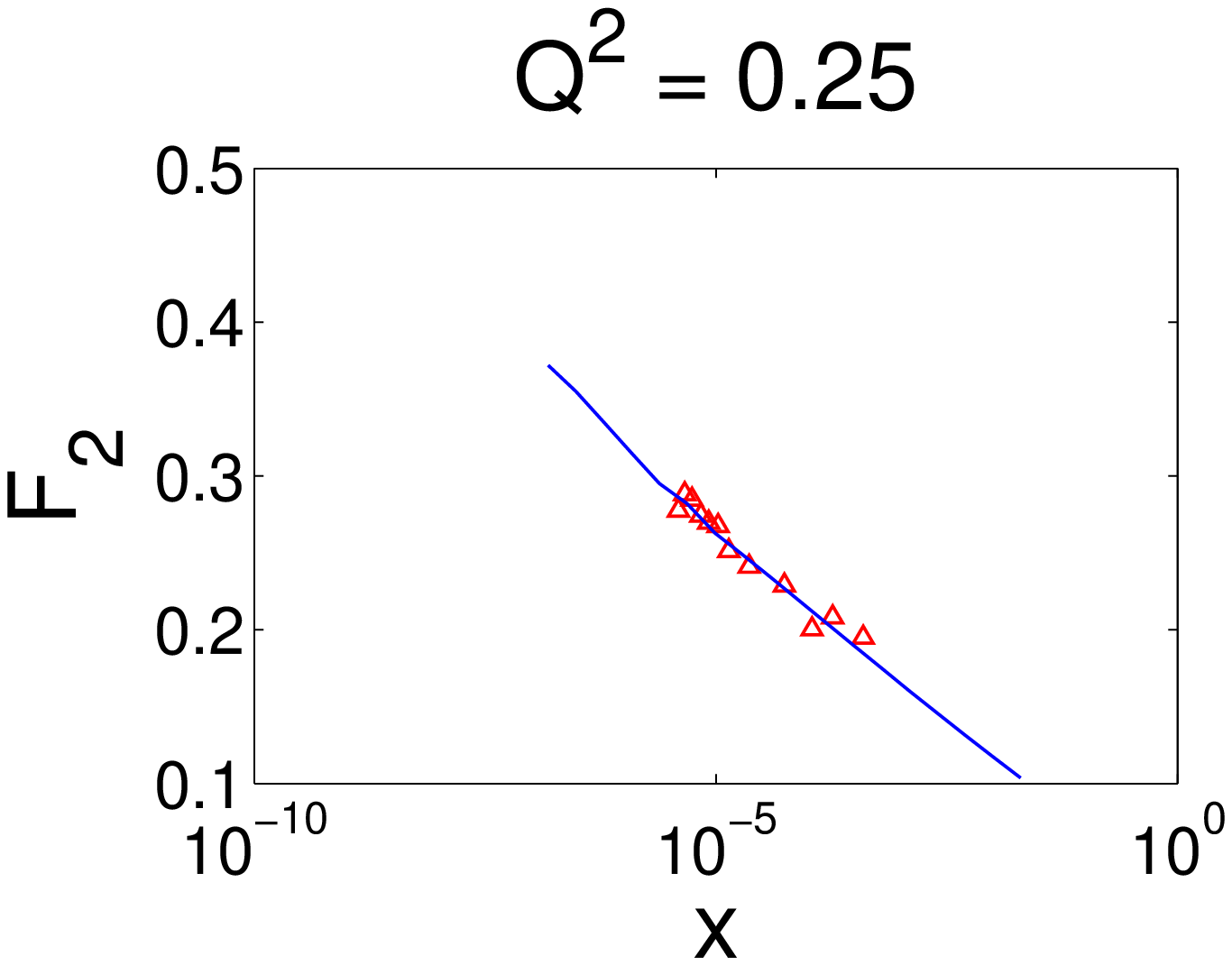,width=50mm,height=30mm}
&\epsfig{file=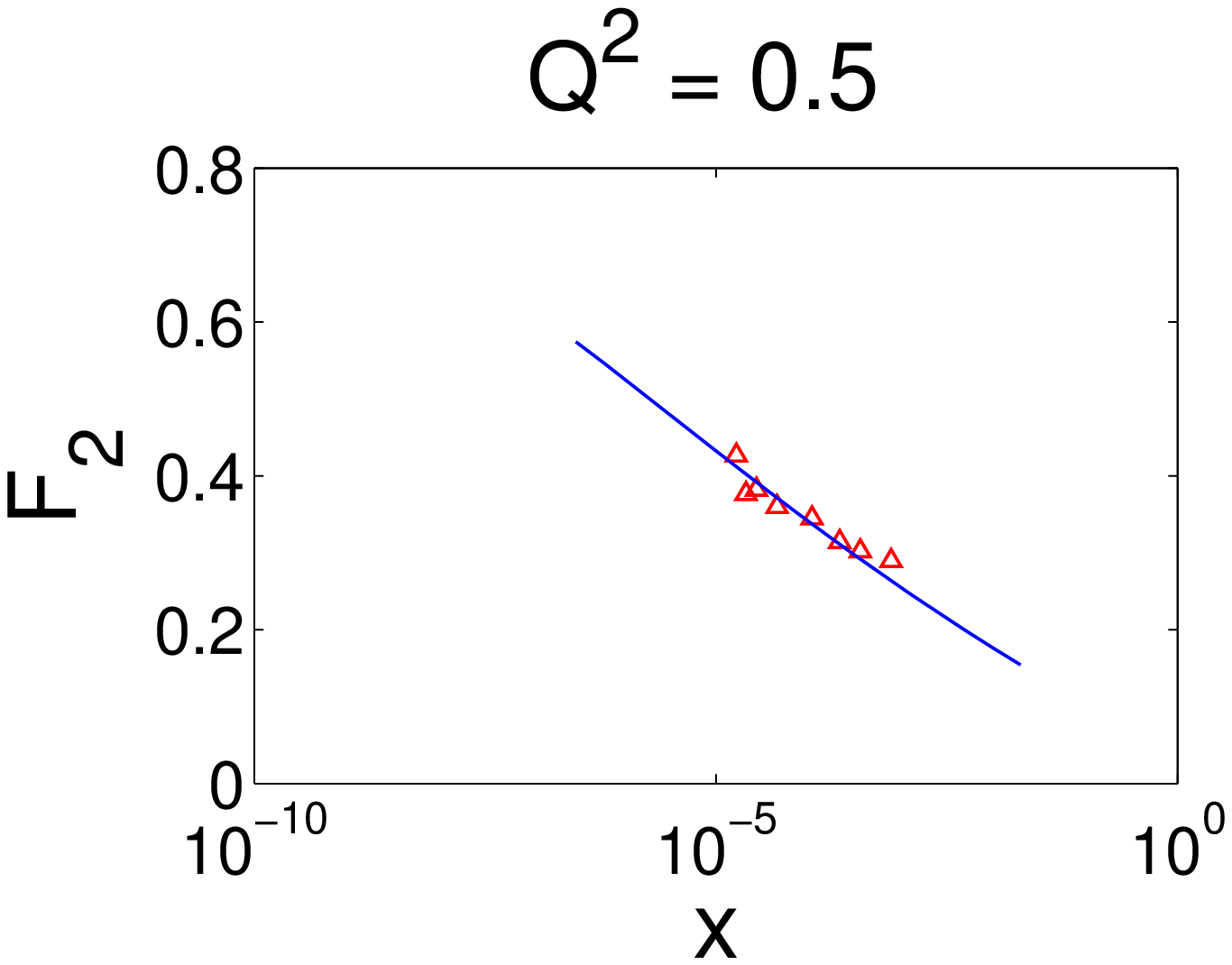,width=50mm,height=30mm}&
\epsfig{file= 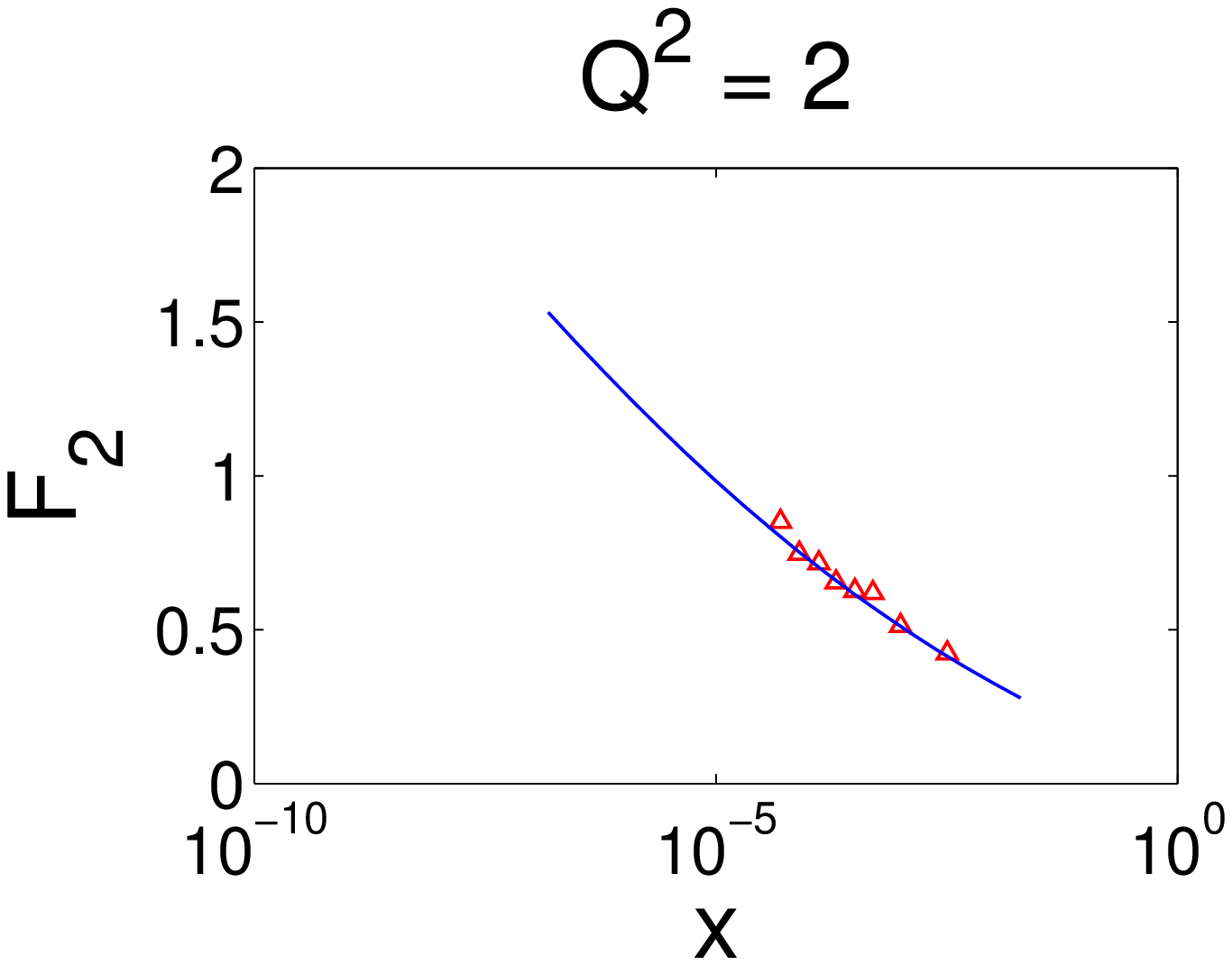,width=50mm,height=30mm} \\
\epsfig{file= 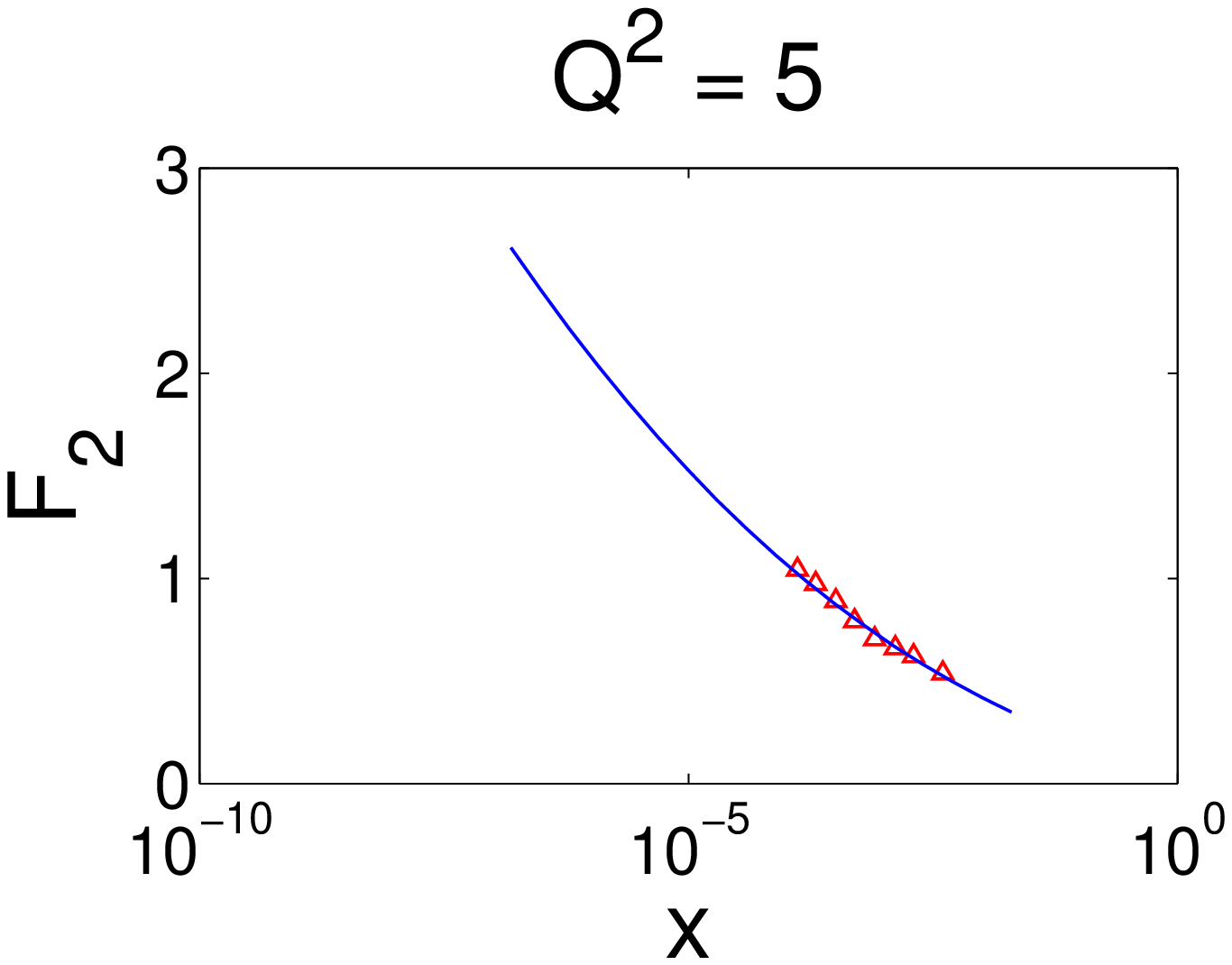,width=50mm,height=30mm}
& \epsfig{file=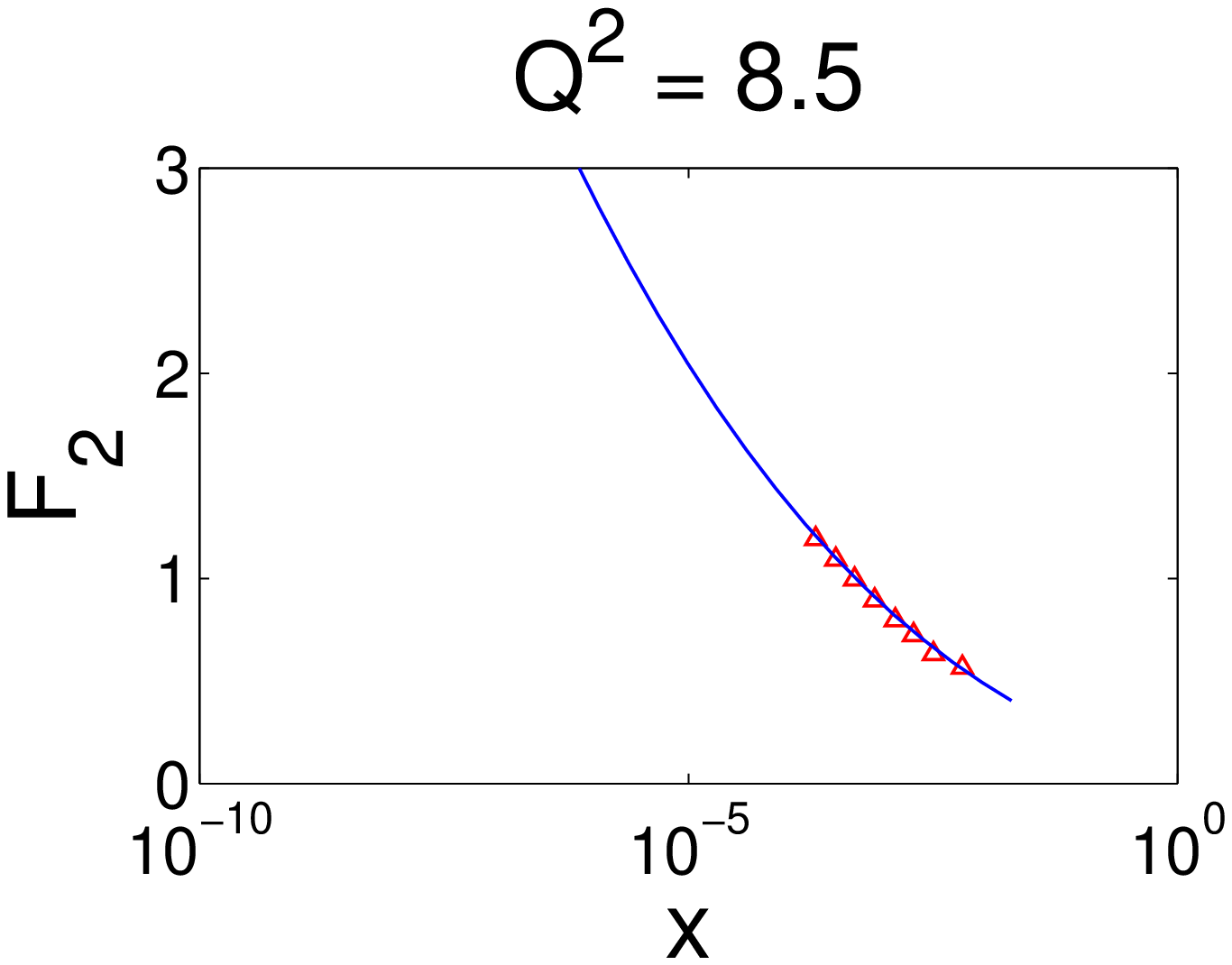,width=50mm,height=30mm}&
\epsfig{file= 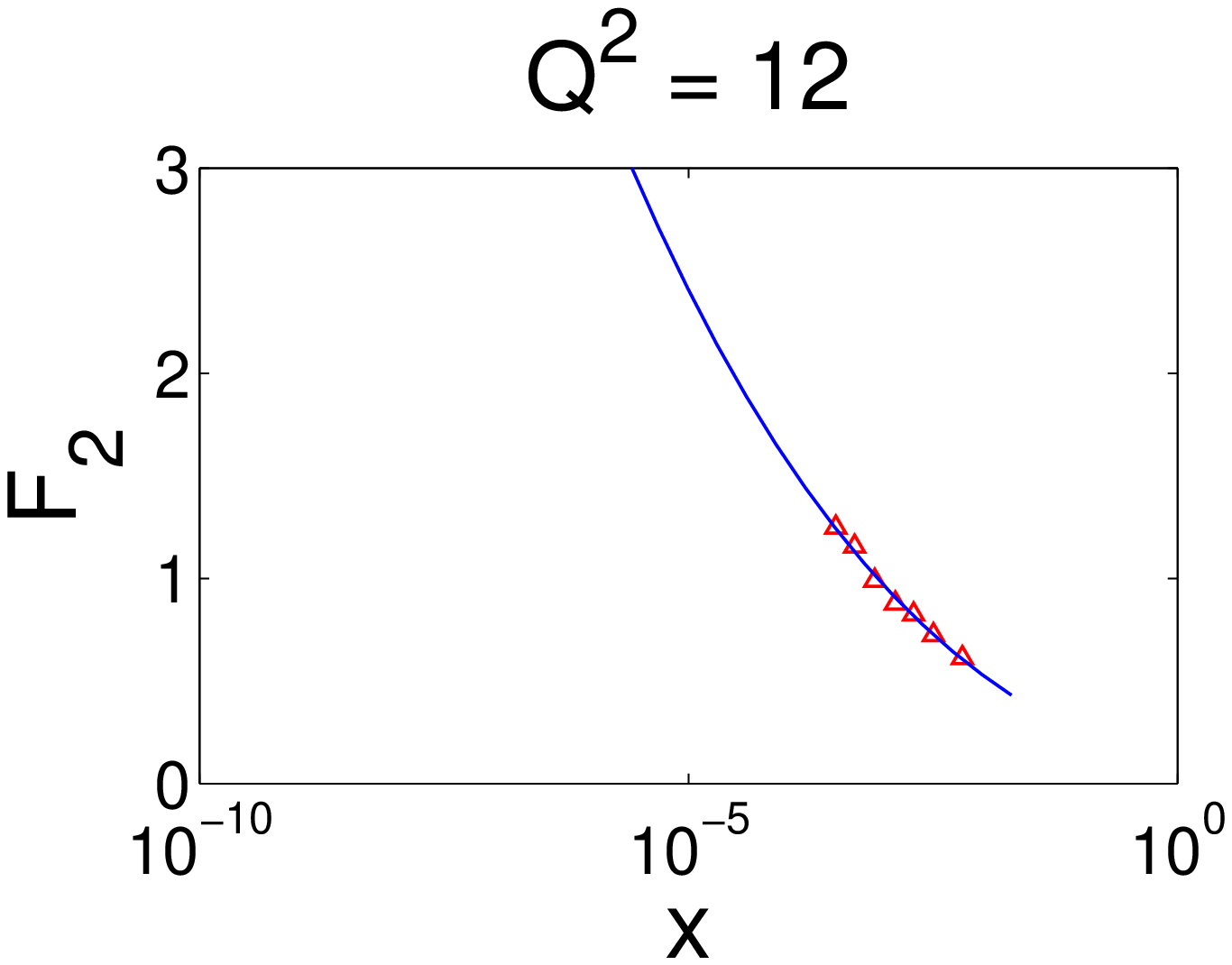,width=50mm,height=30mm} \\
\epsfig{file= 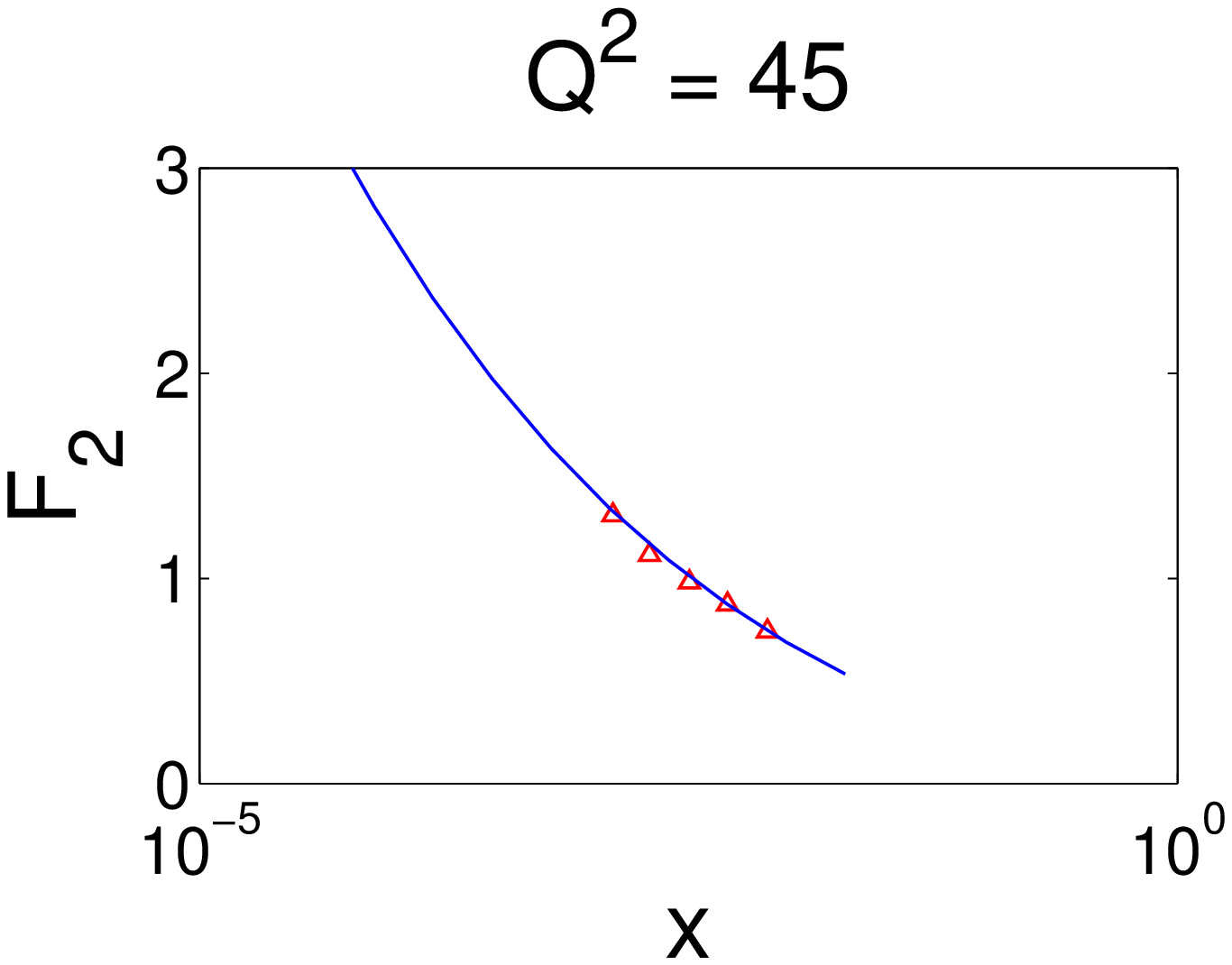,width=50mm,height=30mm}
&\epsfig{file=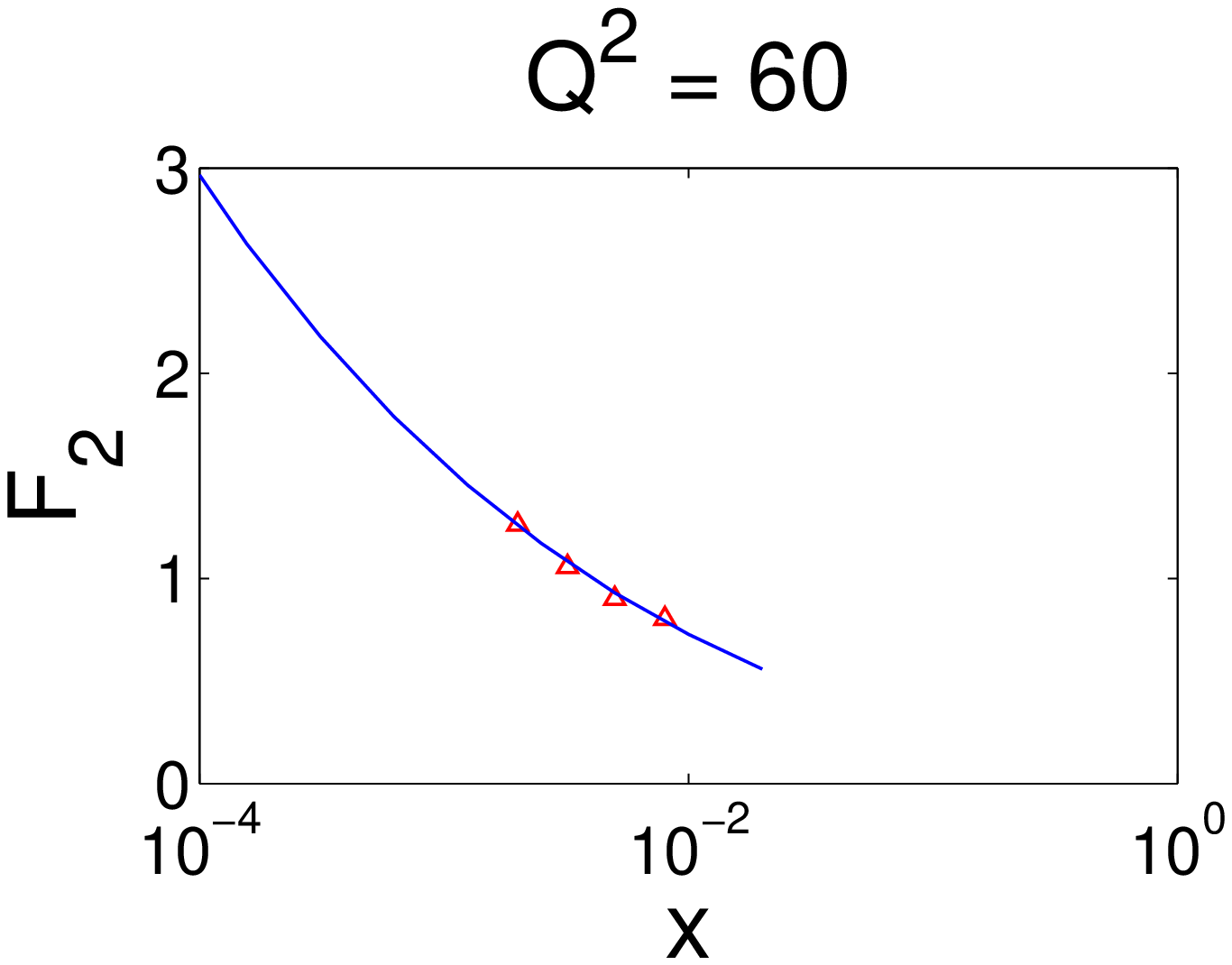,width=50mm,height=30mm}&
\epsfig{file= 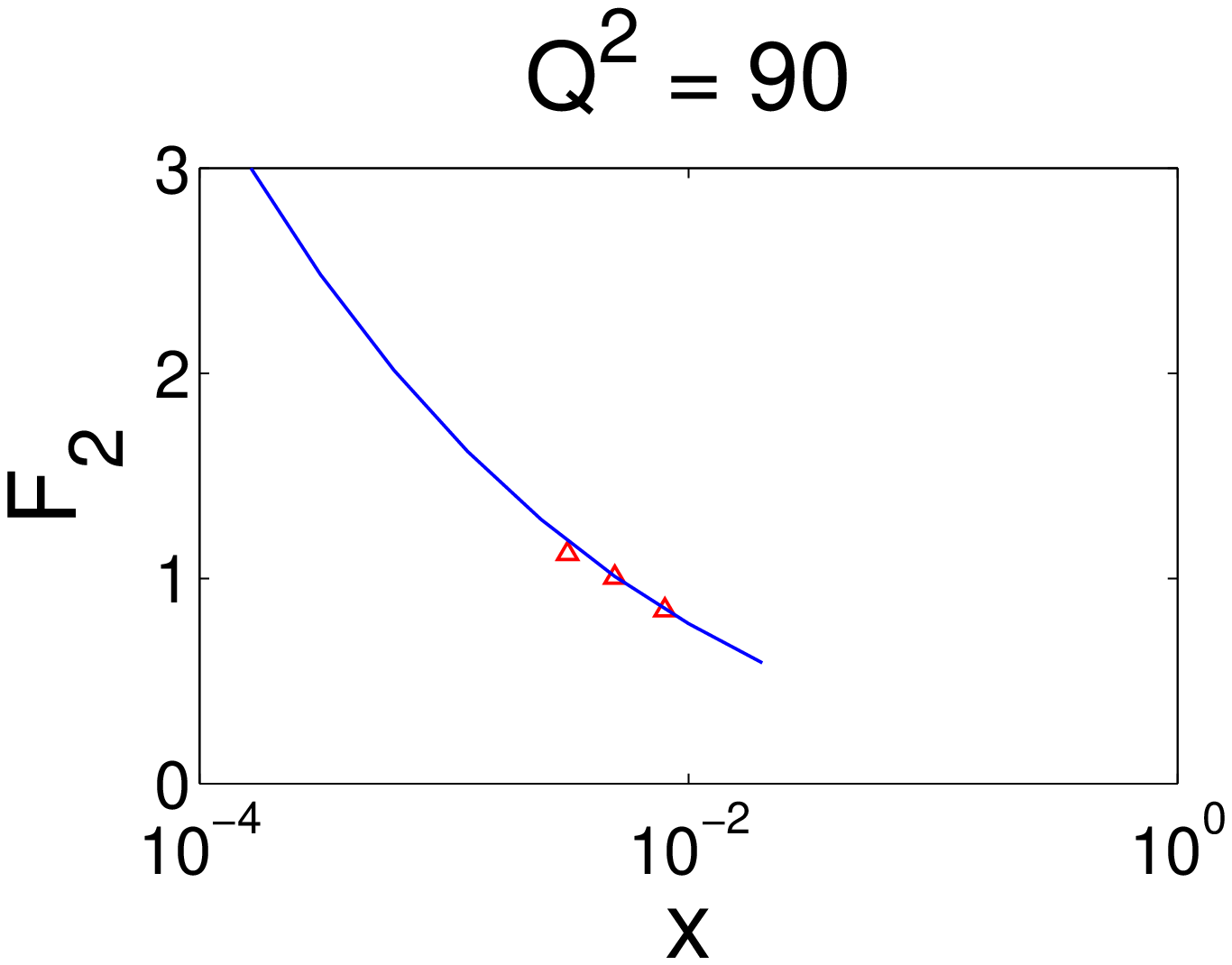,width=50mm,height=30mm}\\
\end{tabular}
\caption{Examples of the fit with \protect\eq{SPJ1}.}
\label{F2}}
%%%%%%%%%%%%%%%%%%%%%%%%%%%%%%%%%%%%%%%%%%%%%%%%%%%%%%%%%%%%%%%%%%%%%%

Using \eq{SPJ2} we can write the formula for \fig{3pjdia},
\bea  \label{SPJ5}
&&\sigma\Lb \gamma^* + p \longrightarrow J/\Psi + M (\fig{3pjdia}) \Rb \,
=\,G_{3P}\,( M^2)^{\Delta_P}\,\int\,d^2\,b\,\exp\Lb -\frac{b^2}{R^2}\Rb\,\times
\\&&
\left|\int dz\,d^2\,r\,\Psi_{\gamma^*}(r,z,Q^2) A(r,x_P)\Psi_{J/\Psi}(r,z)
\right|\,\times\, \left|\int d z'\, d^2\, r' \Psi_{\gamma^*}(r,'z',Q^2)
A(r',x_P)\Psi_{J/\Psi}(r',z')\right|, \nonumber
\eea
where
\beq \label{SPJA}
A\Lb r,x_P;b\Rb\,=\,\int\,d^2b\,\Omega\Lb Y;r,b\Rb.
\eeq
In \eq{SPJ5} we assume that i) that the hard Pomeron in the upper legs of
\fig{3pjdia} does not depend on the impact parameter.
ii) The triple Pomeron vertex ($HP-P-HP$) coupling is the same as the
triple soft Pomeron vertex ($P-P-P$) coupling.
$HP$ denotes the hard Pomeron, while $P$ stands for the soft Pomeron.
$R$ is the radius of the soft interaction and it was taken to be equal
to $R^2\,=\,\Lb 12\,+\,\ln(M^2/s_0)\Rb GeV^{-2}$, with $s_0 = 1GeV^2$.

The product of the wave functions is taken as\cite{GLMPSI}
\bea
\lefteqn{\Psi_{J/\Psi}(r,z=\frac{1}{2})\,\times\,\Psi_{\gamma^*,T}(r;Q^2)\,=\,
\frac{K_F}{48\alpha_{em}}\sqrt{\frac{3\Gamma_{ee}M_{\psi}}{\pi}}\,
\exp\Lb - \,\frac{r^2_{\perp}\,m^2_c}{3\,v^2} \Rb
\,\times} \nonumber\\
& &
\hspace{3cm}\left\{\frac{a^2}{m_c}\,
\left( \zeta\,K_1(\zeta) - \frac{\zeta^2}{4}K_2(\zeta) \right)
+m_c\,\left( \frac{\zeta^2}{2}\,K_2(\zeta) - \zeta\,K_1(\zeta) \right)
\right\},
\label{SPJWFT}
\eea
\bea
&&
\Psi_{J/\Psi}(r,z=\frac{1}{2})\,\times\,\Psi_{\gamma^*,L}(r; Q^2)\,=\,
\frac{K_F}{48\alpha_{em}}\sqrt{\frac{3\Gamma_{ee}M_{\psi}}{\pi}}\,
\exp\Lb - \,\frac{r^2_{\perp}\,m^2_c}{3\,v^2} \Rb
\,\times \nonumber\\
& &
\hspace{3cm}\left\{\frac{Q}{2}
\left( \frac{\zeta^2}{2}\,K_2(\zeta) - \zeta\,K_1(\zeta) \right) \right\}.
\label{SPJWFL}
\eea
$\zeta=a\,r$, $K_i$ (where i=1,2) are the modified Bessel functions,
$\Gamma_{ee}=5.26\,\mbox{KeV}$ is the leptonic width of $J/\Psi$.
\beq \label{SPJAA}
a^2\,=\,z\,( 1 - z)\,Q^2\,+\,m^2_c, \,\,\,\,\,\,\,\mbox{and}
\,\,\,\,\,\,\,\,\,\,\,\,x_P\,\,=\,\,\frac{Q^2\,\,+\,\,M^2}{s}.
\eeq
The exponential factor in \eq{SPJWFT} and \eq{SPJWFL} which  describes the
wave function of $J/\Psi$ meson with velocity $v$,
has been discussed in Ref.\cite{MLR} and references therein.

For the diagram of \fig{3pspjdia} we can write the expression that takes
into account the possible rescatterings before the interaction that
produces a $J/\Psi$ meson,
\bea  \label{SPJ6}
&&\sigma\Lb \gamma^* + p \longrightarrow J/\Psi  + M (\fig{3pspjdia}) \Rb
\,\,\,
=\,\,\,G_{3P}\,( M^2)^{\Delta_P} \,\,\int\,d^2 b\,\, \,\exp\Lb
-\frac{b^2}{R^2}\Rb\,\times \\
&&
\left|\int d z\, d^2\, r \Psi_{\gamma^*}(r,z,Q^2)\,\, A(r,x_P)\,\,
\Lb \,1\,\,\,-\,\,\,N(Y=\ln(1/x);r,b)\,\Rb\,\,\Psi_{J/\Psi}(r,z)\right|\,\times\,
\nonumber \\
&&
\left|\int d z'\, d^2\, r' \Psi_{\gamma^*}(r,'z',Q^2) \,\,A(r',x_P)\,\,\Lb
\,1\,\,\,
-\,\,\,N(Y=\ln(1/x);r',b)\,\Rb\,\,\Psi_{J/\Psi}(r',z')\right|. \nonumber
\eea

The survival probability is the ratio of these two equations (\eq{SPJ5} and
\eq{SPJ6})
\beq \label{SPJ7}
S^2_{3P}(J/\Psi)\,=\,\frac{\sigma\Lb \gamma^* + p \longrightarrow J/\Psi
+ M (\fig{3pspjdia};\,\eq{SPJ6}) \Rb}{
\sigma\Lb \gamma^* + p \longrightarrow J/\Psi + M (\fig{3pjdia};\,\eq{SPJ5})\,
\Rb}\,.
\eeq

The results of our calculations using \eq{SPJ7} are plotted in \fig{spgpsi}.
%%%%%%%%%%%%%%%%%%%%%%%%%%%%%%%%%%%%%%%%%%%%%%%%%%%%%%%%%%%%%%%%%%%%%%%%%
\FIGURE[ht]{
\centerline{\epsfig{file= 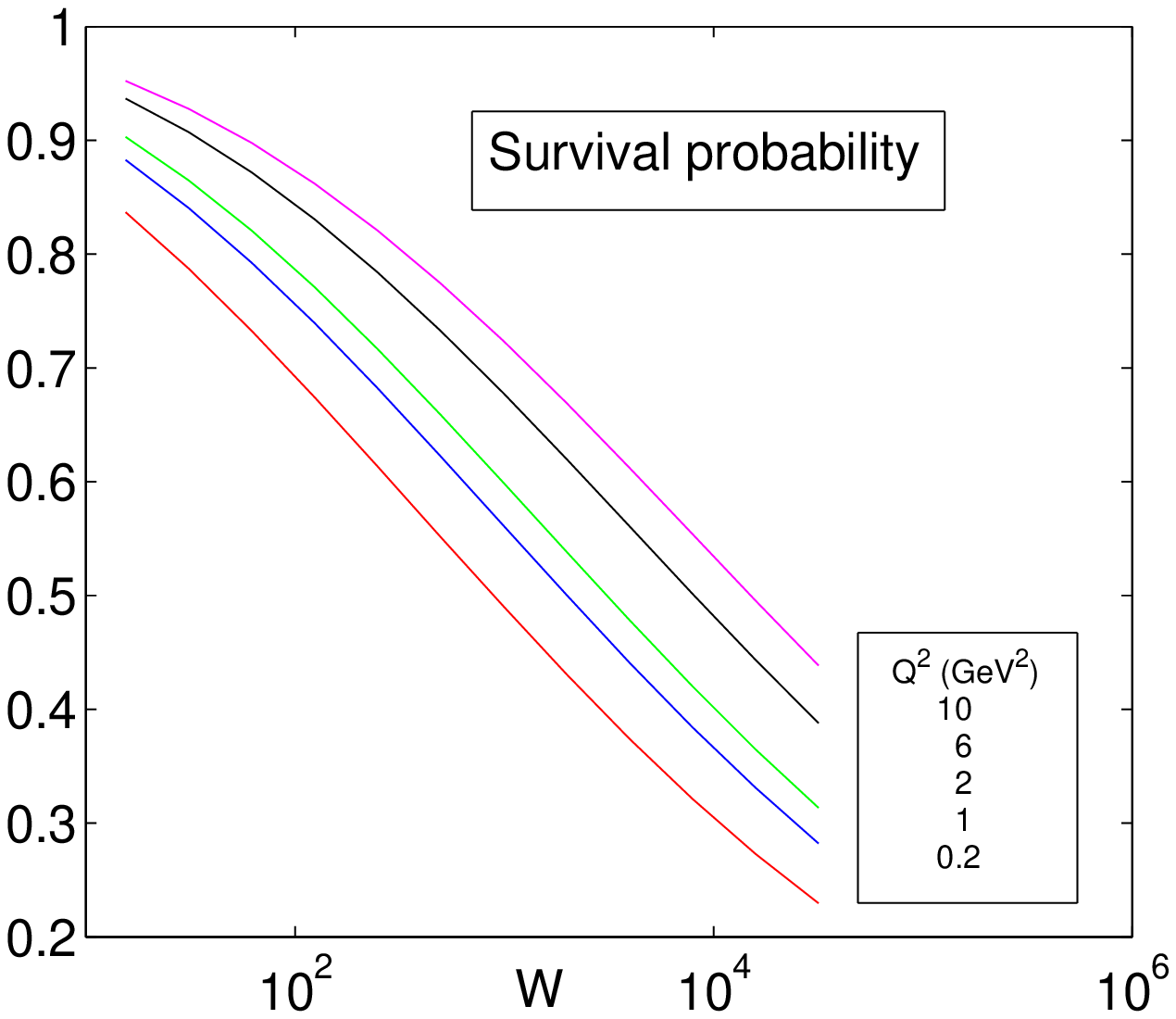,width=90mm,height=90mm}}
\caption{The survival probability for the triple Pomeron vertex in the
reaction of \eq{REACT}.}
\label{spgpsi}}
%%%%%%%%%%%%%%%%%%%%%%%%%%%%%%%%%%%%%%%%%%%%%%%%%%%%%%%%%%%%%%%%%%%%%%%%%%
One can see that $S^2_{3P}$ is smaller than 1, not compatible with the
assumption of Ref.\cite{DG3P}.
This should be taken into account when attempting
to extract the value of the triple Pomeron vertex from the measurement
of the cross section of reaction of \eq{REACT}.

Taking the survival probabilities corrections into account when extracting
the value of $G_{3P}$, it is interesting to study the ratio
$R\,=\,R_p/R_{J/\Psi}$, where
\beq  \label{RATIO1}
R_p \,=\,\frac{\sigma\Lb p + p \to p + M (M \gg m_p) \Rb}
{\sigma\Lb p + p \to p + p\Rb}\,,
\eeq
\beq  \label{RATIO2}
R_{J/\Psi} \,=\,\frac{\sigma \Lb \gamma^* + p \to J/\Psi + M (M \gg m_p) \Rb}
{G_{3P}\Lb \gamma^* + p \to J/\Psi + p \Rb}\,,
\eeq
so as to validate the theoretical estimates, as well as checking the sensitivity
of $G_{3P}$ to the hardness of the coupled Pomerons.
%%%%%%%%%%%%%%%%%%%%%%%%%%%%%%%%%%%%%%%%%%%%%%%%%%%%%%%%%%%%%%%
\section{Conclusions}
%%%%%%%%%%%%%%%%%%%%%%%%%%%%%%%%%%%%%%%%%%%%%%%%%%%%%%%%%%%%%%
In this paper we confirm
the wide spread expectation that the survival
probability for the triple Pomeron vertex is very small\cite{3POLD,3PSC,DG3P}.
However, whereas we find the results of our
two amplitude Model A almost identical to
the results obtained in Ref.\cite{DG3P} (which is also a two amplitude model),
our three amplitude model estimates are an order of magnitude smaller. This
may also influence the $S^2$ calculations for other channels. In
particular, it was noted\cite{heralhc} that the various two amplitude
calculations
of $S^2_{CD}$, relevant to exclusive central Higgs production, are
 in remarkable agreement.
 This optimistic evaluation should be carefully re-examined.

We stress the importance of $J/\Psi$ SD photoproduction (\eq{REACT}).
Comparing its high mass diffraction data with the corresponding $p$-$p$
scattering,
and correcting both channels with their corresponding survival probabilities,
we hope to evaluate both the value of $G_{3P}$ and its dependence on the
Pomeron's hardness.

In this context, we emphasize that even though $S^2_{3P}$ obtained for \eq{REACT}
is considerably higher than $S^2_{3P}$ obtained for $p$-$p$ scattering,
its value is less than unity and should not be neglected.

The details of our three amplitude model B, including calculations for hard
diffraction, and other LRG configurations will be published in a forthcoming
paper.
%%%%%%%%%%%%%%%%%%%%%%%%%%%%%%%%%%%%%%%%%%%%%%%%%%%%%%%%%%%%%%%%%%%%%%%%
\section*{Acknowledgments:}
We are very grateful to Jeremy Miller, Eran Naftali and Alex Prygarin
for fruitful discussions on the subject.
This research was supported in part by the Israel Science Foundation,
founded by the Israeli Academy of Science and Humanities,
by BSF grant \# 20004019 and by a
grant from Israel Ministry of Science, Culture and Sport,
and the Foundation for Basic Research of the Russian Federation.
%%%%%%%%%%%%%%%%%%%%%%%%%%%%%%%%%%%%%%%%%%%%%%%%%%%%%%%%%%%%%%%%%%%%%%%

\end{document}